%% file: article_job_revised.tex
\documentclass[12pt]{iopart} 
\usepackage{epsfig} 
 
\newcommand{\be}{\begin{equation}} 
\newcommand{\ee}{\end{equation}} 
\newcommand{\al}{\alpha} 
\newcommand{\ba}{\beta} 
\newcommand{\ga}{\gamma} 
\newcommand{\de}{\delta} 
\newcommand{\ex}[1]{\rme^{#1}}

\newcommand{\suml}[2]{\sum\limits_{#1}^{#2}} 
 
\newcommand{\ld}{\left.} 
\newcommand{\rd}{\right.} 
\newcommand{\bra}{\left<} 
\newcommand{\ket}{\right>} 
\newcommand{\lb}{\left|} 
\newcommand{\rb}{\right|}

\begin{document} 
\title[$s$-ordered phase-sum and phase-difference distributions] 
{$s$-ordered phase-sum and phase-difference distributions of entangled coherent states} 
\author{A Alexandrescu\footnote[3]{To whom correspondence should be 
    addressed (aalexandrescu@tehfi.pub.ro). Now with Faculty of Electronics and Telecommunications - Optoelectronics Research Center, "Politehnica" University of Bucharest, RO-061071 Bucharest, Romania.} and D H Schiller} 
\address{Department of Physics, University of Siegen, D-57068 Siegen, Germany} 
 
\begin{abstract} 
The $s$-ordered phase-sum and phase-difference distributions are considered 
for Bell-like superpositions of two-mode coherent states. The distributions are sensitive, respectively, 
to the sum and difference of the phases of the entangled coherent states. They show loss of information 
about the entangled state and may take on negative values for some orderings $s$. 
\end{abstract} 
 
\pacs{42.50 Dv, 03.65 Ud}

\section{Introduction} 
 
The quantum phase issue is still an open problem, since there are several nonequivalent proposals to extend the notion of a phase from classical to quantum physics: construction of a phase operator canonically conjugate to the photon number operator \cite{susskind_glogower,pegg_barnett_phase}, derivation of a phase distribution by integrating a phase space distribution over the ``radial'' variables \cite{bandilla_paul,shapiro_wagner} or operational definition by a certain experimental set-up \cite{noh_mandel}. 
Here we choose the phase space approach in order to discuss the phase 
properties of entangled two-mode coherent states. 
The two-mode (or bipartite) states have acquired a great interest due to their potential applications in secure communications \cite{bouwmeester}, quantum teleportation \cite{bouwmeester} as well as in questioning the fundamentals of quantum theory, e.g. in testing Bell inequalities \cite{reid_walls}. 
On the other hand, the $s$-ordered multimode quasi-probability distributions can be sampled point-by-point using unbalanced homodyning \cite{banaszek}. The corresponding phase distributions considered here then follow as marginal distributions. 
 
This paper is organized as follows. In Section \ref{section_charac_prob_func} we calculate for the quasi-Bell state considered the $s$-ordered characteristic function and quasi-probability distribution. The corresponding $s$-ordered phase-sum and phase-difference distributions are derived and discussed in Section \ref{section_phase_diff_distrib}. The concluding Section \ref{summary} contains a brief summary.

\section{Characteristic functions and quasi-probability distributions} 
\label{section_charac_prob_func} 
 
We consider two bosonic modes with annihilation (creation) operators $a$, $b$ ($a^\dagger$, $b^\dagger$). 
One-mode coherent states $\lb\al\ket$ ($\lb\ba\ket$) are defined as eigenstates of $a$ ($b$) with complex 
eigenvalues  $\al=|\al|\exp(\rmi\varphi_\al)$ ($\ba=|\ba|\exp(\rmi\varphi_\ba)$). Denoting by 
$\lb\al,\ba\ket=\lb\al\ket\otimes\lb\ba\ket$ the two-mode coherent states, we consider quasi-Bell states 
of the form 
\begin{equation} 
\lb\Psi\ket={\cal N}\:(\mu\lb\al,\ba\ket+\nu\lb-\al,-\ba\ket), 
\label{psi_state} 
\end{equation} 
where $\mu$ and $\nu$ are complex parameters restricted by $|\mu|^2+|\nu|^2=1$ and $\cal N$ is the 
normalization constant 
\be 
{\cal N}=\left\{1+2\mbox{Re}(\mu \nu^\ast)\exp[-2(|\al|^2+|\ba|^2)] \right\}^{-\frac{1}2}.\label{N_const}\\ 
\ee 
The superposition (\ref{psi_state}) represents a general two-mode Schr\" odinger cat state. Particular states obtained for $(\mu,\nu)=(1,\pm 1)/\sqrt{2}$ and 
$(\mu,\nu)=(1,\pm \rmi)/\sqrt{2}$ are referred to as even/odd Schr\" odinger cat and 
Yurke-Stoler \cite{yurke_stoler} states, respectively. By a suitable choice of the parameters $\al$, $\ba$, $\mu$ and $\nu$ the state $\lb\Psi\ket$ may exhibit maximal bipartite entanglement \cite{wang} or two-mode squeezing \cite{chai}. 
 
Given the density operator $\rho$ of a system one defines the complex-valued  $s$-ordered characteristic function by \cite{cahill_glauber, barnett_radmore}: 
\be 
\chi(\xi,\eta;s)=\exp\left(\frac{s}{2}(|\xi|^2+|\eta|^2)\right)\Tr\{\rho D(\xi,\eta)\}, 
\label{chi_p_definition} 
\ee 
where $D(\xi,\eta)\equiv\exp(\xi a^\dagger-\xi^\ast a)\exp(\eta b^\dagger-\eta^\ast b)$ is 
the two-mode displacement operator with complex displacements $\xi$ and $\eta$, and $s$ is a complex 
parameter. The discrete values $s=1,0,-1$ 
correspond to normal, symmetric and antinormal ordering of the boson operators, respectively.  
The characteristic function of the state (\ref{psi_state}) with density operator 
$\rho = \lb\Psi\ket\bra\Psi\rb$ is easily obtained as 
\begin{eqnarray} 
\fl\chi(\xi,\eta;s)={\cal N}^2\exp\left(-\frac{1-s}{2}\left(|\xi|^2+|\eta|^2\right)\right)\left\{\: |\mu|^2 
 \exp(\xi\al^\ast-\xi^\ast\al+\eta\ba^\ast-\eta^\ast\ba) \rd\nonumber\\ 
\lo+ |\nu|^2\exp(-\xi\al^\ast+\xi^\ast\al-\eta\ba^\ast+\eta^\ast\ba) \nonumber\\ 
\lo + \exp(-2|\al|^2-2|\ba|^2)\ [\: \mu^\ast\nu\exp(\xi\al^\ast+\xi^\ast\al+\eta\ba^\ast+\eta^\ast\ba)\nonumber\\ 
\lo+\ld\!\mu\nu^\ast\exp(-\xi\al^\ast-\xi^\ast\al-\eta\ba^\ast-\eta^\ast\ba)] 
\right\}. 
\label{chi_psi} 
\end{eqnarray} 
The $s$-ordered quasi-probability distribution can be derived from the characteristic function by taking its Fourier transform  
\be 
\fl W(\ga,\de;s)=\frac{1}{\pi^4}\int\rmd^2\xi\int\rmd^2\eta\,\chi(\xi,\eta;s)\exp(\ga\xi^\ast-\ga^\ast\xi) 
 \exp(\de\eta^\ast-\de^\ast\eta). 
\label{pseudo_prob_distrib} 
\ee 
In order for this transform to exist one must have $\mbox{Re}(s)<1$. For $s=-1$ one obtains the (positive valued) $Q$(or Husimi)-function, for $s=0$ the Wigner function and for $s=1$ the (highly singular) Glauber-Sudarshan $P$-representation. 
Inserting the characteristic function (\ref{chi_psi}) in relation (\ref{pseudo_prob_distrib}) one obtains the following expression for the $s$-ordered quasi-probability distribution of the state (\ref{psi_state}): 
\begin{eqnarray} 
\fl W(\ga,\de;s)=\frac{4{\cal N}^2}{\pi^2(1-s)^2}\exp\left(-\frac{2}{1-s}(|\al|^2+|\ba|^2)\right) 
 \exp\left(-\frac{2}{1-s}(|\ga|^2+|\de|^2)\right)\nonumber\\ 
\!\!\!\!\!\!\!\!\!\!\!\!\!\!\!\!\!\!\! 
\lo\times \!\!\left\{|\mu|^2\exp\!\left(\!\frac{2(\al^\ast\ga+\al\ga^\ast+\ba^\ast\de+\ba\de^\ast)}{1-s}\right) 
\!+|\nu|^2\exp\!\left(\!-\frac{2(\al^\ast\ga+\al\ga^\ast+\ba^\ast\de+\ba\de^\ast)}{1-s}\right)\!\!\rd\nonumber\\ 
\!\!\!\!\!\!\!\!\!\!\!\!\!\!\!\!\!\!\! 
\lo+ \!\exp\left(\frac{2(1+s)}{1-s}(|\al|^2+|\ba|^2)\right)\left[ 
\mu^\ast\nu\exp\!\left(\!\frac{2(\al^\ast\ga-\al\ga^\ast+\ba^\ast\de-\ba\de^\ast)}{1-s}\right)\rd\nonumber\\ 
\!\!\!\!\!\!\!\!\!\!\!\!\!\!\!\!\!\!\! 
\lo+\ld\!\ld\!\mu\nu^\ast\exp\!\left(\!-\frac{2(\al^\ast\ga-\al\ga^\ast+\ba^\ast\de-\ba\de^\ast)}{1-s}\right)\right] 
\right\}. 
\label{quasi_prob_psi} 
\end{eqnarray} 
It has the property $[W(\ga,\de;s)]^\ast=W(\ga,\de;s^\ast)$ and thus is real-valued only for real values of $s$. In this case considered henceforth 
the distribution $W(\ga,\de;s)$ has two major contributions: the first one containing $|\mu|^2$, $|\nu|^2$ and exponentials of real numbers is always positive (Gaussian terms), whereas the second one containing $\mu^\ast\nu$, $\mu\nu^\ast$ 
and exponentials of imaginary numbers may take on negative values as well (interference terms). 
By varying the value of the parameter $s$ one can enhance or suppress the 
interference terms in comparison with the Gaussian ones. This results in always positive distributions $W(\ga,\de;s)$ only for $s \leq -1$. If $-1<s<1$, the distributions take on negative values as well.

\section{Phase-sum and phase-difference distributions} 
\label{section_phase_diff_distrib} 
 
In experiments one measures only relative phases, e.g. with respect to the phase of a local oscillator as in homodyne detection \cite{bachor, leonhardt}. Therefore, it makes sense to address the problem of phase-difference and, complementary, of phase-sum distributions in the phase operator approach \cite{barnett_pegg} as well as in the phase space approach 
\cite{freyberger_schleich}. 
 
The main steps in deriving the phase-sum and phase-difference distributions from the 
two-mode quasi-probability distribution (\ref{pseudo_prob_distrib}) are: 
\begin{enumerate} 
\item The normalization condition of $W(\ga,\de;s)$ is written in polar coordinates $(\ga=|\ga| \exp(\rmi\varphi_\ga)$, $\de=|\de| \exp(\rmi\varphi_\de))$: 
\be 
\int_0^\infty\rmd|\ga|\int_0^\infty\rmd|\de|\int_0^{2\pi}\rmd\varphi_\ga 
 \int_0^{2\pi}\rmd\varphi_\de\:|\ga||\de|\:W(\ga,\de;s)=1. 
\label{W_polar_norm} 
\ee 
\item The phase variables $\varphi_\ga$ and $\varphi_\de$ are replaced by the phase-sum and 
-difference: 
\be 
\phi_+=\varphi_\de+\varphi_\ga,\quad\phi_-=\varphi_\de-\varphi_\ga. 
\label{phase_plus_minus_def} 
\ee 
In going from two one-mode phases defined on $2\pi$-domains to their sum and difference a problem occurs because $\phi_+$ and $\phi_-$ have $4\pi$-ranges according to (\ref{phase_plus_minus_def}). This is not compatible with measurements that cannot tell a phase value $\phi$ from $\phi\pm 2\pi$. To cast the phase-sum and -difference into $2\pi$-ranges, we follow reference \cite{sanchez} and replace $W(\ga,\de;s)$ by 
\be 
\fl {\cal W}(|\ga|,|\de|,\phi_+,\phi_-;s)\!\!=\!\!\frac{1}{2}\left[W(|\ga|,|\de|,\varphi_\ga,\varphi_\de;s) 
  \!+\!W(|\ga|,|\de|,\varphi_\ga+\pi,\varphi_\de+\pi;s)\right] 
\label{W_transformation} 
\ee 
with $\varphi_\ga$ and $\varphi_\de$ on the right-hand side substituted in terms of $\phi_\pm$ from relation 
(\ref{phase_plus_minus_def}). 
The quasi-probability distribution ${\cal W}$ is then normalized according to 
\be 
\fl \int_0^\infty\rmd|\ga|\int_0^\infty\rmd|\de|\int_0^{2\pi}\rmd\phi_+\int_0^{2\pi}\rmd\phi_-\:|\ga||\de| 
\:{\cal W}(|\ga|,|\de|,\phi_+,\phi_-;s)=1. 
\label{cal_W_polar_norm} 
\ee 
\item The phase-sum (phase-difference) distribution is obtained by integration over the 
``radial'' variables $|\ga|$, $|\de|$ and over $\phi_-$ ($\phi_+$): 
\be 
\fl {\cal P}^{(\pm)}(\phi_\pm;s)=\int_0^\infty\rmd|\ga|\int_0^\infty\rmd|\de| 
 \int_0^{2\pi}\rmd\phi_\mp\:|\ga||\de|\:{\cal W}(|\ga|,|\de|,\phi_+,\phi_-;s). 
\label{phase_diff_def} 
\ee 
\end{enumerate} 
Applying the above procedure to the quasi-probability distribution (\ref{quasi_prob_psi}), we first expand the exponential terms of the distribution (\ref{W_transformation}) into series of Bessel functions \cite{prudnikov}, 
then we integrate over the appropriate phase variable ($\phi_-$ or $\phi_+$) and finally over the ``radial'' variables \cite{prudnikov}. 
We write the final result for the $s$-ordered phase-sum and phase-difference distributions in the form of a genuine phase distribution: 
\be 
{\cal P}^{(\pm)}(\phi_\pm;s)=\frac{1}{2\pi}\left[1+2\suml{n=1}{\infty}c_n^{(\pm)}(s) 
\cos n\left(\phi_\pm-\phi^\prime_\pm \right)\right], 
\label{phase_diff_prob} 
\ee 
with state-dependent phases $\phi^\prime_\pm \equiv \varphi_\ba\pm\varphi_\al$ and  coefficients $c_n^{(\pm)}(s)$ given by: 
\begin{eqnarray} 
c_n^{(\pm)}(s) = \ &{\cal N}^2& \: \frac{\pi}{2}\:\left\{ 
  \:I_n^{(+)}(\case{|\al|^2}{1-s})\:I_n^{(+)}(\case{|\ba|^2}{1-s})\rd\nonumber\\ 
\!\!\!\!\!\!\!\!\!\!\!\!\!\! 
&&\!\!\lo+ \ld\!\!(\pm)^n\: 2 \:\mbox{Re}(\mu \nu^\ast) \: \ex{-2(|\al|^2+|\ba|^2)} 
 \:I_n^{(-)}(\case{|\al|^2}{1-s})\:I_n^{(-)}(\case{|\ba|^2}{1-s})\right\}. 
 \label{cn_bessel} 
\end{eqnarray} 
Here $\cal N$ is the normalization constant (\ref{N_const}) and we have introduced the notations 
\begin{eqnarray} 
I_n^{(\pm)}(x) &\equiv & \sqrt{x} \: \ex{\mp x} \left( I_{\case{n-1}2}(x)\pm I_{\case{n+1}2}(x)\right) \label{cn_bessel_2} \\ 
&=&\sqrt{\frac{2}{\pi}}\ \frac{\Gamma(\frac{n}{2}+1)}{\Gamma(n+1)}\: 
(2x)^\frac{n}{2}\:\ex{\mp 2x} M({\textstyle \frac{n}{2}}+1,n+1,\pm 2x)\label{cn_kummer} , 
\end{eqnarray} 
where $I_\nu(x)$ is the modified Bessel function of order $\nu$ and $M(a,b,x)$ is  Kummer's confluent hypergeometric function. The relation (\ref{cn_kummer}) follows from (\ref{cn_bessel_2}) by relating $I_\nu(x)$ to $M(a,b,x)$ and using then one of the recurrence relations \cite{prudnikov}. 
When calculating the phase-sum and -difference distributions (\ref{phase_diff_def}) with $\phi_\pm$ spanning a $2\pi$-domain, part of the information about the state 
$\lb\Psi\ket$ is lost. This loss of information is due to the imposed $2\pi$-periodicity in $\phi_{\pm}$ which renders the transformation (\ref{phase_plus_minus_def}) nonbijective \cite{sanchez}. In our case the superposition (\ref{psi_state}) depends on two complex numbers $\mu$ and $\nu$ subject to the condition $|\mu|^2+|\nu|^2=1$, so that only three real independent parameters are left. Correspondingly the density operator and therefore the quasi-probability distribution (\ref{quasi_prob_psi}) depend on the three parameters $|\mu|^2-|\nu|^2$, $\mbox{Re}(\mu \nu^\ast)$ and $\mbox{Im}(\mu \nu^\ast)$. In comparison, the phase distributions (\ref{phase_diff_prob}) depend only on  $\mbox{Re}(\mu \nu^\ast)$. The terms containing $|\mu|^2-|\nu|^2$ and $\mbox{Im}(\mu \nu^\ast)$ were lost already at the level of the quasi-probability 
distribution $\cal W$ when evaluated according to relation (\ref{W_transformation}). 
An additional loss of information occurs after performing the integration (\ref{phase_diff_def}) and is related with the dependence on the state-related phases $\varphi_\al$ and $\varphi_\ba$: they enter the distribution 
${\cal P}^{(+)}$ (${\cal P}^{(-)}$) only through the corresponding combination $\varphi_\ba+\varphi_\al$ ($\varphi_\ba-\varphi_\al$). The right-hand side of (\ref{phase_diff_prob}) is a Fourier series expansion with the sine terms missing. This makes the distributions symmetric about the origine $\phi_\pm = \phi^\prime_\pm$. 
 
We note that less information is lost when considering the marginal one-mode phase distributions. Thus, integrating $W(\ga,\de;s)$ in relation (\ref{quasi_prob_psi}) over the remaining variables, we obtain the phase distribution for mode 1 in the form 
\begin{eqnarray} 
\fl {\cal P}^{(1)}(\varphi_\ga;s)= 
\frac{1}{2\pi}\left\{1+2\sum_{k=1}^\infty 
\left[c^{(1)}_{2k}(s)\:\cos(2k(\varphi_\ga-\varphi_\al))\rd\rd\nonumber\\ 
\!\!\!\!\!\!\!\!+\ld\ld c^{(1)}_{2k-1}(s)\cos((2k-1)(\varphi_\ga-\varphi_\al)) 
+ d^{(1)}_{2k-1}(s)\sin((2k-1)(\varphi_\ga-\varphi_\al))\right] \right\}, 
\label{one-mode_phase_distribution} 
\end{eqnarray} 
where the coefficients are given by 
\begin{eqnarray} 
c^{(1)}_{2k}(s) &=& {\cal N}^2\:\sqrt{\frac{\pi}{2}}\: 
\{\: I_{2k}^{(+)}(\case{|\al|^2}{1-s})+  2 \:\mbox{Re}(\mu \nu^\ast)\: \ex{-2(|\al|^2+|\ba|^2)} \: I_{2k}^{(-)}(\case{|\al|^2}{1-s})\:\} , \nonumber \\ 
c^{(1)}_{2k-1}(s) &=& {\cal N}^2\:\sqrt{\frac{\pi}{2}}\: 
\{\:(|\mu|^2-|\nu|^2) \: I_{2k-1}^{(+)}(\case{|\al|^2}{1-s})\:\} , \nonumber \\ 
d^{(1)}_{2k-1}(s) &=& {\cal N}^2\:\sqrt{\frac{\pi}{2}}\: 
\{\:2 \:\mbox{Im}(\mu \nu^\ast) \: \ex{-2(|\al|^2+|\ba|^2)} \: I_{2k-1}^{(-)}(\case{|\al|^2}{1-s})\:\} . 
\label{one-mode_phase_coeff} 
\end{eqnarray} 
The marginal phase distribution ${\cal P}^{(2)}(\varphi_\de;s)$ of mode 2 has the same form (\ref{one-mode_phase_distribution}) in terms of the phase variable $\varphi_\de-\varphi_\ba$ and corresponding coefficients $c^{(2)}_{k}(s)$ and $d^{(2)}_{k}(s)$. The latter are obtained formally by interchanging 
$|\al| \leftrightarrow |\ba|$ in (\ref{one-mode_phase_coeff}). 
These one-mode phase distributions depend on all but one of the relevant parameters defining the state (\ref{psi_state}). Thus, the distribution of mode 1 (2) does not depend on the phase $\varphi_\ba$ ($\varphi_\al$) of the other mode. Due to the sine terms present in (\ref{one-mode_phase_distribution}), the distributions are no longer symmetric about the phase origine. Note also that the sine terms with even indices are still missing. 
 
In Figure \ref{wigner_phase_distrib} we display phase-difference and in Figure \ref{wigner_phase_sum_distrib} phase-sum distributions for even (upper subfigures (a), (b)) and odd (lower subfigures (c), (d)) Schr\"odinger cat states. Subfigures (a) and (c) on the left show the phase distributions (\ref{phase_diff_prob}) over the plane $(|\al|^2,\phi_\pm - \phi^\prime_\pm)$ for $s=0$ and $|\al|=|\ba|$. The dependence on the ordering parameter $s$ and the phases $\phi_\pm - \phi^\prime_\pm$ is plotted for three values of $s$ (-1, 0 and 0.4) and for $|\al|=|\ba|=1$ in  subfigures (b) and (d) on the right. The phase distributions are symmetric about $\phi_\pm = \phi^\prime_\pm = \varphi_\ba\pm\varphi_\al$, with a peak or dip at the symmetry point. For small values of $|\al|$ they approach the uniform distribution as long as the normalization constant $\cal N$ in (\ref{cn_bessel}) stays finite at $|\al| = 0$. However, if $1+2\mbox{Re}(\mu \nu^\ast) = 0$, as in the example of Figure 1(c), a non-uniform distribution is approached. Regarding the dependence on the ordering parameter, the phase distributions corresponding to $s>-1$ may take on negative values and may develop strong oscillations, especially for $s>0$. This is due to the fact that the interference terms in (\ref{quasi_prob_psi}) become more important and may overwhelm the Gaussian terms. In general, the smaller the value of the ordering parameter is, the smoother is the corresponding phase distribution. Note that quasi-probability distributions $W$ taking on also negative values may nevertheless yield positive-valued phase distributions $\cal P$, as can be seen from Figures 1(b) and 2(d) for $s=0$. This is not surprising, as the marginal distributions of a quasi-distribution may be true distributions.

\begin{figure}[b] 
\begin{center} 
\scriptsize 
\hspace*{-3cm}\input{wigner_even_state.tex}\hspace*{-2cm}\input{p_even_articol.tex} 
  
\vspace*{-2.5cm} 
\hspace*{-3cm}\input{wigner_odd_state.tex}\hspace*{-2cm}\input{p_odd_articol.tex} 
\normalsize 
\end{center} 
\vspace*{-3cm} 
\caption{Phase-difference distributions. Left:  From the Wigner function ($s=0$) for 
(a) even and (c) odd Schr\" odinger cat states with $|\al|=|\ba|$. 
Right: Dependence on the ordering parameter, $s=-1$ (\full), $s=0$ (\broken) 
and $s=0.4$ (\dashed) for (b) even and (d) odd Schr\" odinger cat states with $|\al|=|\ba|=1$.} 
\label{wigner_phase_distrib} 
\end{figure}
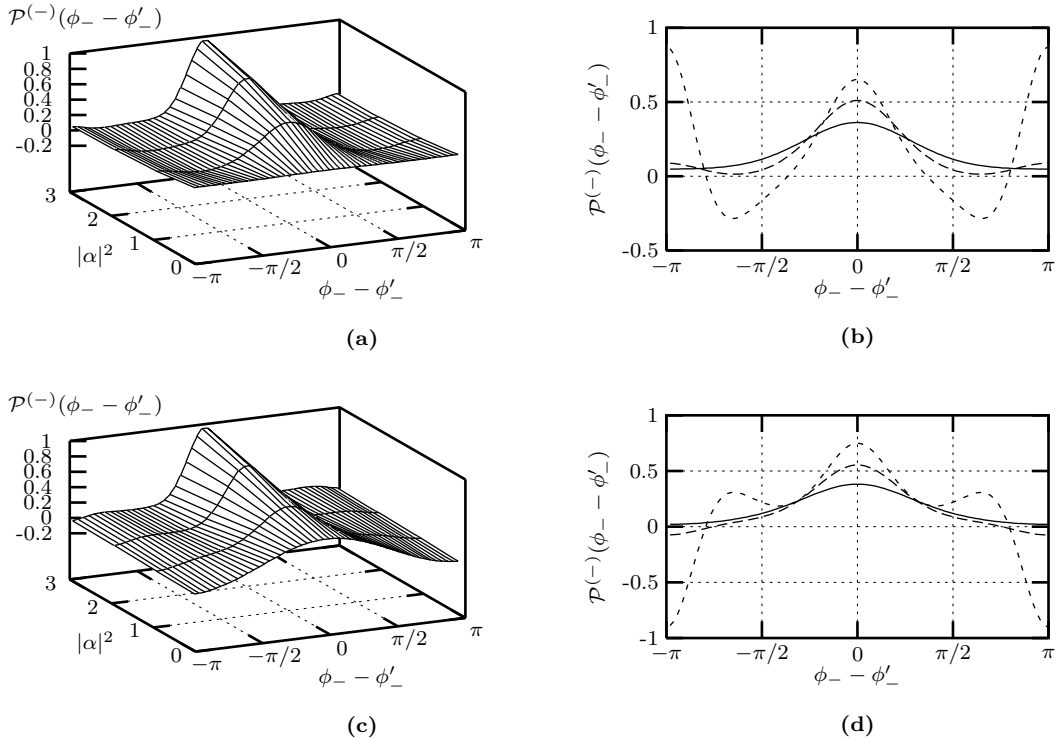 
\begin{figure}[t] 
\begin{center} 
\scriptsize 
\hspace*{-3cm}\input{pw_sum_even.tex}\hspace*{-2cm}\input{pp_sum_even.tex} 
  
\vspace*{-2.5cm} 
\hspace*{-3cm}\input{pw_sum_odd.tex}\hspace*{-2cm}\input{pp_sum_odd.tex} 
\normalsize 
\end{center} 
\vspace*{-3.5cm} 
\caption{Phase-sum distributions for the same input data as in Figure \ref{wigner_phase_distrib}.} 
\label{wigner_phase_sum_distrib} 
\end{figure}
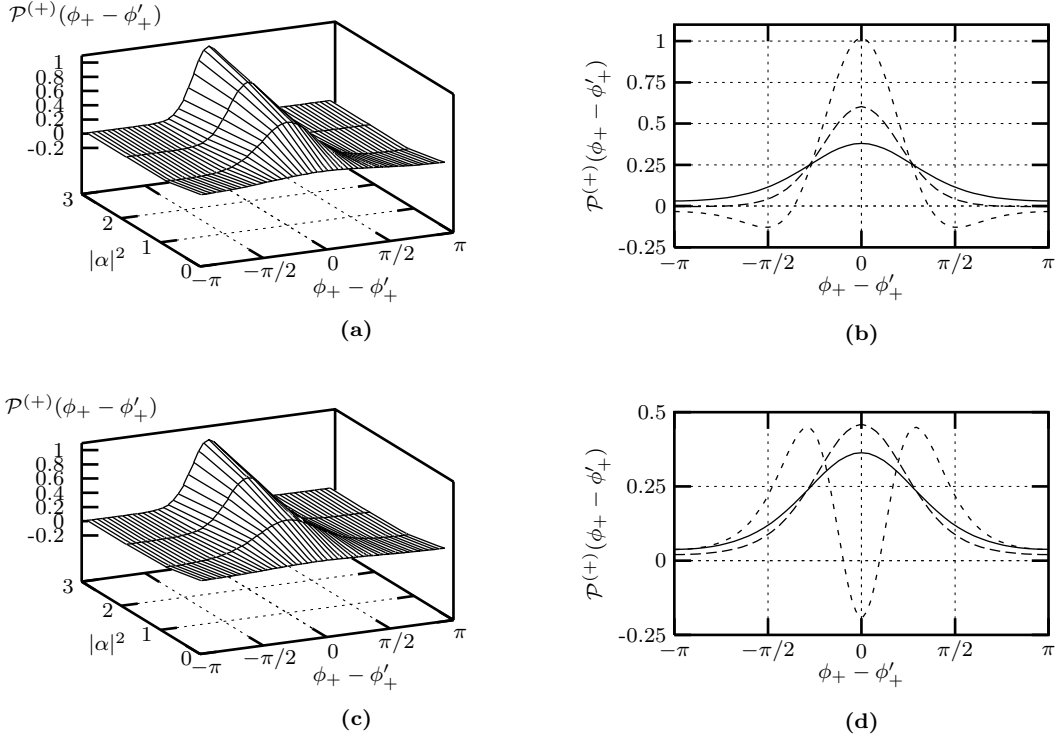 
Given the quasi-probability distribution (\ref{phase_diff_prob}) one can calculate the expectation value of various phase-dependent functions. 
Here we give the central trigonometric moments (we drop the indices $\pm$ on $\phi$, $\phi^\prime$ and $c_n$): 
\be 
\bra\cos n(\phi-\phi^\prime)\ket=c_n,\qquad 
\bra\sin n(\phi-\phi^\prime)\ket=0, 
\label{trig_moments} 
\ee 
the variances squared 
\be 
\fl (\Delta\cos n(\phi-\phi^\prime))^2=\frac{1}{2}\left(1-2c_n^2+c_{2n}\right), 
\qquad (\Delta\sin n(\phi-\phi^\prime))^2=\frac{1}{2}(1-c_{2n}) 
\ee 
and the mean value and variance squared of the phase $\phi$ for the $2\pi$-window $[\phi_0-\pi,\phi_0+\pi]$: 
\begin{eqnarray} 
\bra\phi\ket=\phi_0+2\sum_{n=1}^\infty \frac{(-)^n}{n}c_n\sin n(\phi_0-\phi^\prime),\\ 
(\Delta\phi)^2=\frac{\pi^2}{3}-(\bra\phi\ket-\phi_0)^2+4\sum_{n=1}^\infty 
 \frac{(-)^n}{n^2}c_n\cos n(\phi_0-\phi^\prime). 
\end{eqnarray} 
The latter expressions simplify considerably by choosing $\phi_0=\phi^\prime$ for the center of the window, i.e. 
$\phi_0 = \phi^\prime_\pm = \varphi_\beta\pm \varphi_\alpha$ for the phase-sum/phase-difference distributions. It follows from (\ref{trig_moments}) that the coefficients $c_n(s)$ are the expectation values of $\cos n(\phi-\phi^\prime)$ and therefore must satisfy the bound $|c_n(s)|<1$ for a genuine phase distribution. This bound may be violated, however, for quasi-distributions taking on negative values. This is the reason why such phase distributions, as in Figure 2(d), may develop a (negative) dip instead of a (positive) peak at $\phi=\phi^\prime$. Similar considerations apply to the coefficients $c^{(i)}_{n}(s)$ and $d^{(i)}_{n}(s)$, $i=1,2$, of the one-mode phase distributions (\ref{one-mode_phase_distribution}), which are the expectation values of the corresponding cosine and sine functions multiplying them.

\section{Summary} 
\label{summary} 
 
In this paper we have considered various marginal phase distributions obtained from the $s$-ordered two-mode quasi-probability distribution for the entangled two-mode coherent state in (\ref{psi_state}). Proper non-negative distributions are obtained for $s \leq -1$ and improper or quasi-distributions for $-1< s < 1$. The latter take on negative values as well and may yield unphysical expectation values for the (central) trigonometric moments. The Glauber-Sudarshan $P$-representation corresponding to $s=1$ is not defined. 
 
As expected and contrary to the quasi-probability distribution $W(\ga,\de;s)$ in (\ref{quasi_prob_psi}), the marginal phase distributions are not sensitive to the full set of parameters characterizing the state (\ref{psi_state}). Thus, the phase distributions for one mode depend only on its own phase and do not feel the phase of the other mode. Similarly, the phase-sum and phase-difference distributions involve, respectively, only the sum $\varphi_\beta + \varphi_\alpha$ and difference $\varphi_\beta - \varphi_\alpha$ of the two state-related phases. In addition, the two latter distributions do not depend on $|\mu|^2 - |\nu|^2$ and $\mbox{Im}(\mu \nu^\ast)$, but only on $\mbox{Re}(\mu \nu^\ast)$, as a consequence of casting the $4\pi$-periodic distribution $W$ (considered as a function of $\phi_+$ and $\phi_-$) into a $2\pi$-periodic one $\cal W$ according to relation (\ref{W_transformation}). Regarded as Fourier series expansions, the phase-sum and phase-difference distributions have only (even) cosine terms and are therefore symmetric about the origine. The one-mode phase distributions, however, are unsymmetric due to the (odd) sine terms.

\end{document}

%% file: wigner_even_state.tex
\begingroup%
  \makeatletter%
  \newcommand{\GNUPLOTspecial}{%
    \@sanitize\catcode`\%=14\relax\special}%
  \setlength{\unitlength}{0.1bp}%
\begin{picture}(3600,2160)(0,-500)%
\special{psfile=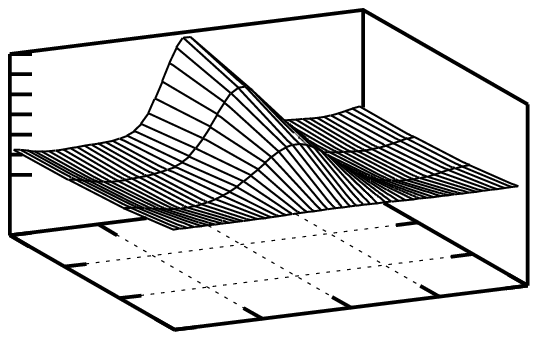 llx=0 lly=0 urx=720 ury=504 rwi=7200}
\put(1114,1594){\makebox(0,0){${\cal P}^{(-)}(\phi_--\phi^\prime_-)$}}%
\put(988,1463){\makebox(0,0)[r]{1}}%
\put(988,1405){\makebox(0,0)[r]{0.8}}%
\put(988,1347){\makebox(0,0)[r]{0.6}}%
\put(988,1289){\makebox(0,0)[r]{0.4}}%
\put(988,1231){\makebox(0,0)[r]{0.2}}%
\put(988,1173){\makebox(0,0)[r]{0}}%
\put(988,1115){\makebox(0,0)[r]{-0.2}}%
\put(2155,564){\makebox(0,0){$\phi_--\phi^\prime_-$}}%
\put(2155,384){\makebox(0,0){\bf (a)}}%
\put(1571,619){\makebox(0,0){$-\pi$}}%
\put(1825,650){\makebox(0,0){$-\pi/2$}}%
\put(2080,682){\makebox(0,0){0}}%
\put(2334,714){\makebox(0,0){$\pi/2$}}%
\put(2588,746){\makebox(0,0){$\pi$}}%
\put(1150,700){\makebox(0,0){$|\al|^2$}}%
\put(1004,928){\makebox(0,0)[r]{3}}%
\put(1162,838){\makebox(0,0)[r]{2}}%
\put(1320,747){\makebox(0,0)[r]{1}}%
\put(1478,656){\makebox(0,0)[r]{0}}%
\end{picture}%
\endgroup
 

%% file: p_even_articol.tex
\begingroup%
  \makeatletter%
  \newcommand{\GNUPLOTspecial}{%
    \@sanitize\catcode`\%=14\relax\special}%
  \setlength{\unitlength}{0.1bp}%
{\GNUPLOTspecial{!
/gnudict 256 dict def
gnudict begin
/Color false def
/Solid false def
/gnulinewidth 5.000 def
/userlinewidth gnulinewidth def
/vshift -20 def
/dl {10 mul} def
/hpt_ 31.5 def
/vpt_ 31.5 def
/hpt hpt_ def
/vpt vpt_ def
/M {moveto} bind def
/L {lineto} bind def
/R {rmoveto} bind def
/V {rlineto} bind def
/vpt2 vpt 2 mul def
/hpt2 hpt 2 mul def
/Lshow { currentpoint stroke M
  0 vshift R show } def
/Rshow { currentpoint stroke M
  dup stringwidth pop neg vshift R show } def
/Cshow { currentpoint stroke M
  dup stringwidth pop -2 div vshift R show } def
/UP { dup vpt_ mul /vpt exch def hpt_ mul /hpt exch def
  /hpt2 hpt 2 mul def /vpt2 vpt 2 mul def } def
/DL { Color {setrgbcolor Solid {pop []} if 0 setdash }
 {pop pop pop Solid {pop []} if 0 setdash} ifelse } def
/BL { stroke userlinewidth 2 mul setlinewidth } def
/AL { stroke userlinewidth 2 div setlinewidth } def
/UL { dup gnulinewidth mul /userlinewidth exch def
      10 mul /udl exch def } def
/PL { stroke userlinewidth setlinewidth } def
/LTb { BL [] 0 0 0 DL } def
/LTa { AL [1 udl mul 2 udl mul] 0 setdash 0 0 0 setrgbcolor } def
/LT0 { PL [] 1 0 0 DL } def
/LT1 { PL [4 dl 2 dl] 0 1 0 DL } def
/LT2 { PL [2 dl 3 dl] 0 0 1 DL } def
/LT3 { PL [1 dl 1.5 dl] 1 0 1 DL } def
/LT4 { PL [5 dl 2 dl 1 dl 2 dl] 0 1 1 DL } def
/LT5 { PL [4 dl 3 dl 1 dl 3 dl] 1 1 0 DL } def
/LT6 { PL [2 dl 2 dl 2 dl 4 dl] 0 0 0 DL } def
/LT7 { PL [2 dl 2 dl 2 dl 2 dl 2 dl 4 dl] 1 0.3 0 DL } def
/LT8 { PL [2 dl 2 dl 2 dl 2 dl 2 dl 2 dl 2 dl 4 dl] 0.5 0.5 0.5 DL } def
/Pnt { stroke [] 0 setdash
   gsave 1 setlinecap M 0 0 V stroke grestore } def
/Dia { stroke [] 0 setdash 2 copy vpt add M
  hpt neg vpt neg V hpt vpt neg V
  hpt vpt V hpt neg vpt V closepath stroke
  Pnt } def
/Pls { stroke [] 0 setdash vpt sub M 0 vpt2 V
  currentpoint stroke M
  hpt neg vpt neg R hpt2 0 V stroke
  } def
/Box { stroke [] 0 setdash 2 copy exch hpt sub exch vpt add M
  0 vpt2 neg V hpt2 0 V 0 vpt2 V
  hpt2 neg 0 V closepath stroke
  Pnt } def
/Crs { stroke [] 0 setdash exch hpt sub exch vpt add M
  hpt2 vpt2 neg V currentpoint stroke M
  hpt2 neg 0 R hpt2 vpt2 V stroke } def
/TriU { stroke [] 0 setdash 2 copy vpt 1.12 mul add M
  hpt neg vpt -1.62 mul V
  hpt 2 mul 0 V
  hpt neg vpt 1.62 mul V closepath stroke
  Pnt  } def
/Star { 2 copy Pls Crs } def
/BoxF { stroke [] 0 setdash exch hpt sub exch vpt add M
  0 vpt2 neg V  hpt2 0 V  0 vpt2 V
  hpt2 neg 0 V  closepath fill } def
/TriUF { stroke [] 0 setdash vpt 1.12 mul add M
  hpt neg vpt -1.62 mul V
  hpt 2 mul 0 V
  hpt neg vpt 1.62 mul V closepath fill } def
/TriD { stroke [] 0 setdash 2 copy vpt 1.12 mul sub M
  hpt neg vpt 1.62 mul V
  hpt 2 mul 0 V
  hpt neg vpt -1.62 mul V closepath stroke
  Pnt  } def
/TriDF { stroke [] 0 setdash vpt 1.12 mul sub M
  hpt neg vpt 1.62 mul V
  hpt 2 mul 0 V
  hpt neg vpt -1.62 mul V closepath fill} def
/DiaF { stroke [] 0 setdash vpt add M
  hpt neg vpt neg V hpt vpt neg V
  hpt vpt V hpt neg vpt V closepath fill } def
/Pent { stroke [] 0 setdash 2 copy gsave
  translate 0 hpt M 4 {72 rotate 0 hpt L} repeat
  closepath stroke grestore Pnt } def
/PentF { stroke [] 0 setdash gsave
  translate 0 hpt M 4 {72 rotate 0 hpt L} repeat
  closepath fill grestore } def
/Circle { stroke [] 0 setdash 2 copy
  hpt 0 360 arc stroke Pnt } def
/CircleF { stroke [] 0 setdash hpt 0 360 arc fill } def
/C0 { BL [] 0 setdash 2 copy moveto vpt 90 450  arc } bind def
/C1 { BL [] 0 setdash 2 copy        moveto
       2 copy  vpt 0 90 arc closepath fill
               vpt 0 360 arc closepath } bind def
/C2 { BL [] 0 setdash 2 copy moveto
       2 copy  vpt 90 180 arc closepath fill
               vpt 0 360 arc closepath } bind def
/C3 { BL [] 0 setdash 2 copy moveto
       2 copy  vpt 0 180 arc closepath fill
               vpt 0 360 arc closepath } bind def
/C4 { BL [] 0 setdash 2 copy moveto
       2 copy  vpt 180 270 arc closepath fill
               vpt 0 360 arc closepath } bind def
/C5 { BL [] 0 setdash 2 copy moveto
       2 copy  vpt 0 90 arc
       2 copy moveto
       2 copy  vpt 180 270 arc closepath fill
               vpt 0 360 arc } bind def
/C6 { BL [] 0 setdash 2 copy moveto
      2 copy  vpt 90 270 arc closepath fill
              vpt 0 360 arc closepath } bind def
/C7 { BL [] 0 setdash 2 copy moveto
      2 copy  vpt 0 270 arc closepath fill
              vpt 0 360 arc closepath } bind def
/C8 { BL [] 0 setdash 2 copy moveto
      2 copy vpt 270 360 arc closepath fill
              vpt 0 360 arc closepath } bind def
/C9 { BL [] 0 setdash 2 copy moveto
      2 copy  vpt 270 450 arc closepath fill
              vpt 0 360 arc closepath } bind def
/C10 { BL [] 0 setdash 2 copy 2 copy moveto vpt 270 360 arc closepath fill
       2 copy moveto
       2 copy vpt 90 180 arc closepath fill
               vpt 0 360 arc closepath } bind def
/C11 { BL [] 0 setdash 2 copy moveto
       2 copy  vpt 0 180 arc closepath fill
       2 copy moveto
       2 copy  vpt 270 360 arc closepath fill
               vpt 0 360 arc closepath } bind def
/C12 { BL [] 0 setdash 2 copy moveto
       2 copy  vpt 180 360 arc closepath fill
               vpt 0 360 arc closepath } bind def
/C13 { BL [] 0 setdash  2 copy moveto
       2 copy  vpt 0 90 arc closepath fill
       2 copy moveto
       2 copy  vpt 180 360 arc closepath fill
               vpt 0 360 arc closepath } bind def
/C14 { BL [] 0 setdash 2 copy moveto
       2 copy  vpt 90 360 arc closepath fill
               vpt 0 360 arc } bind def
/C15 { BL [] 0 setdash 2 copy vpt 0 360 arc closepath fill
               vpt 0 360 arc closepath } bind def
/Rec   { newpath 4 2 roll moveto 1 index 0 rlineto 0 exch rlineto
       neg 0 rlineto closepath } bind def
/Square { dup Rec } bind def
/Bsquare { vpt sub exch vpt sub exch vpt2 Square } bind def
/S0 { BL [] 0 setdash 2 copy moveto 0 vpt rlineto BL Bsquare } bind def
/S1 { BL [] 0 setdash 2 copy vpt Square fill Bsquare } bind def
/S2 { BL [] 0 setdash 2 copy exch vpt sub exch vpt Square fill Bsquare } bind def
/S3 { BL [] 0 setdash 2 copy exch vpt sub exch vpt2 vpt Rec fill Bsquare } bind def
/S4 { BL [] 0 setdash 2 copy exch vpt sub exch vpt sub vpt Square fill Bsquare } bind def
/S5 { BL [] 0 setdash 2 copy 2 copy vpt Square fill
       exch vpt sub exch vpt sub vpt Square fill Bsquare } bind def
/S6 { BL [] 0 setdash 2 copy exch vpt sub exch vpt sub vpt vpt2 Rec fill Bsquare } bind def
/S7 { BL [] 0 setdash 2 copy exch vpt sub exch vpt sub vpt vpt2 Rec fill
       2 copy vpt Square fill
       Bsquare } bind def
/S8 { BL [] 0 setdash 2 copy vpt sub vpt Square fill Bsquare } bind def
/S9 { BL [] 0 setdash 2 copy vpt sub vpt vpt2 Rec fill Bsquare } bind def
/S10 { BL [] 0 setdash 2 copy vpt sub vpt Square fill 2 copy exch vpt sub exch vpt Square fill
       Bsquare } bind def
/S11 { BL [] 0 setdash 2 copy vpt sub vpt Square fill 2 copy exch vpt sub exch vpt2 vpt Rec fill
       Bsquare } bind def
/S12 { BL [] 0 setdash 2 copy exch vpt sub exch vpt sub vpt2 vpt Rec fill Bsquare } bind def
/S13 { BL [] 0 setdash 2 copy exch vpt sub exch vpt sub vpt2 vpt Rec fill
       2 copy vpt Square fill Bsquare } bind def
/S14 { BL [] 0 setdash 2 copy exch vpt sub exch vpt sub vpt2 vpt Rec fill
       2 copy exch vpt sub exch vpt Square fill Bsquare } bind def
/S15 { BL [] 0 setdash 2 copy Bsquare fill Bsquare } bind def
/D0 { gsave translate 45 rotate 0 0 S0 stroke grestore } bind def
/D1 { gsave translate 45 rotate 0 0 S1 stroke grestore } bind def
/D2 { gsave translate 45 rotate 0 0 S2 stroke grestore } bind def
/D3 { gsave translate 45 rotate 0 0 S3 stroke grestore } bind def
/D4 { gsave translate 45 rotate 0 0 S4 stroke grestore } bind def
/D5 { gsave translate 45 rotate 0 0 S5 stroke grestore } bind def
/D6 { gsave translate 45 rotate 0 0 S6 stroke grestore } bind def
/D7 { gsave translate 45 rotate 0 0 S7 stroke grestore } bind def
/D8 { gsave translate 45 rotate 0 0 S8 stroke grestore } bind def
/D9 { gsave translate 45 rotate 0 0 S9 stroke grestore } bind def
/D10 { gsave translate 45 rotate 0 0 S10 stroke grestore } bind def
/D11 { gsave translate 45 rotate 0 0 S11 stroke grestore } bind def
/D12 { gsave translate 45 rotate 0 0 S12 stroke grestore } bind def
/D13 { gsave translate 45 rotate 0 0 S13 stroke grestore } bind def
/D14 { gsave translate 45 rotate 0 0 S14 stroke grestore } bind def
/D15 { gsave translate 45 rotate 0 0 S15 stroke grestore } bind def
/DiaE { stroke [] 0 setdash vpt add M
  hpt neg vpt neg V hpt vpt neg V
  hpt vpt V hpt neg vpt V closepath stroke } def
/BoxE { stroke [] 0 setdash exch hpt sub exch vpt add M
  0 vpt2 neg V hpt2 0 V 0 vpt2 V
  hpt2 neg 0 V closepath stroke } def
/TriUE { stroke [] 0 setdash vpt 1.12 mul add M
  hpt neg vpt -1.62 mul V
  hpt 2 mul 0 V
  hpt neg vpt 1.62 mul V closepath stroke } def
/TriDE { stroke [] 0 setdash vpt 1.12 mul sub M
  hpt neg vpt 1.62 mul V
  hpt 2 mul 0 V
  hpt neg vpt -1.62 mul V closepath stroke } def
/PentE { stroke [] 0 setdash gsave
  translate 0 hpt M 4 {72 rotate 0 hpt L} repeat
  closepath stroke grestore } def
/CircE { stroke [] 0 setdash 
  hpt 0 360 arc stroke } def
/Opaque { gsave closepath 1 setgray fill grestore 0 setgray closepath } def
/DiaW { stroke [] 0 setdash vpt add M
  hpt neg vpt neg V hpt vpt neg V
  hpt vpt V hpt neg vpt V Opaque stroke } def
/BoxW { stroke [] 0 setdash exch hpt sub exch vpt add M
  0 vpt2 neg V hpt2 0 V 0 vpt2 V
  hpt2 neg 0 V Opaque stroke } def
/TriUW { stroke [] 0 setdash vpt 1.12 mul add M
  hpt neg vpt -1.62 mul V
  hpt 2 mul 0 V
  hpt neg vpt 1.62 mul V Opaque stroke } def
/TriDW { stroke [] 0 setdash vpt 1.12 mul sub M
  hpt neg vpt 1.62 mul V
  hpt 2 mul 0 V
  hpt neg vpt -1.62 mul V Opaque stroke } def
/PentW { stroke [] 0 setdash gsave
  translate 0 hpt M 4 {72 rotate 0 hpt L} repeat
  Opaque stroke grestore } def
/CircW { stroke [] 0 setdash 
  hpt 0 360 arc Opaque stroke } def
/BoxFill { gsave Rec 1 setgray fill grestore } def
/Symbol-Oblique /Symbol findfont [1 0 .167 1 0 0] makefont
dup length dict begin {1 index /FID eq {pop pop} {def} ifelse} forall
currentdict end definefont
end
}}%
\begin{picture}(1800,1080)(0,-1040)%
{\GNUPLOTspecial{"
gnudict begin
gsave
0 0 translate
0.100 0.100 scale
0 setgray
newpath
1.000 UL
LTb
1.000 UL
LTa
270 180 M
1440 0 V
1.000 UL
LTb
270 180 M
63 0 V
1377 0 R
-63 0 V
1.000 UL
LTa
270 460 M
1440 0 V
1.000 UL
LTb
270 460 M
63 0 V
1377 0 R
-63 0 V
1.000 UL
LTa
270 740 M
1440 0 V
1.000 UL
LTb
270 740 M
63 0 V
1377 0 R
-63 0 V
1.000 UL
LTa
270 1020 M
1440 0 V
1.000 UL
LTb
270 1020 M
63 0 V
1377 0 R
-63 0 V
1.000 UL
LTa
270 180 M
0 840 V
1.000 UL
LTb
270 180 M
0 63 V
0 777 R
0 -63 V
1.000 UL
LTa
630 180 M
0 840 V
1.000 UL
LTb
630 180 M
0 63 V
0 777 R
0 -63 V
1.000 UL
LTa
990 180 M
0 840 V
1.000 UL
LTb
990 180 M
0 63 V
0 777 R
0 -63 V
1.000 UL
LTa
1350 180 M
0 840 V
1.000 UL
LTb
1350 180 M
0 63 V
0 777 R
0 -63 V
1.000 UL
LTa
1710 180 M
0 840 V
1.000 UL
LTb
1710 180 M
0 63 V
0 777 R
0 -63 V
1.000 UL
LTb
270 180 M
1440 0 V
0 840 V
-1440 0 V
0 -840 V
1.000 UL
LT0
284 487 M
15 0 V
14 0 V
15 1 V
14 0 V
14 0 V
15 0 V
14 1 V
15 0 V
14 1 V
14 1 V
15 0 V
14 2 V
15 1 V
14 1 V
14 2 V
15 2 V
14 2 V
15 2 V
14 3 V
14 3 V
15 3 V
14 4 V
15 4 V
14 4 V
14 5 V
15 5 V
14 6 V
15 5 V
14 7 V
14 6 V
15 7 V
14 7 V
15 7 V
14 7 V
14 8 V
15 7 V
14 8 V
15 7 V
14 7 V
14 7 V
15 6 V
14 6 V
15 5 V
14 5 V
14 4 V
15 3 V
14 2 V
15 1 V
14 1 V
14 -1 V
15 -1 V
14 -2 V
15 -3 V
14 -4 V
14 -5 V
15 -5 V
14 -6 V
15 -6 V
14 -7 V
14 -7 V
15 -7 V
14 -8 V
15 -7 V
14 -8 V
14 -7 V
15 -7 V
14 -7 V
15 -7 V
14 -6 V
14 -7 V
15 -5 V
14 -6 V
15 -5 V
14 -5 V
14 -4 V
15 -4 V
14 -4 V
15 -3 V
14 -3 V
14 -3 V
15 -2 V
14 -2 V
15 -2 V
14 -2 V
14 -1 V
15 -1 V
14 -2 V
15 0 V
14 -1 V
14 -1 V
15 0 V
14 -1 V
15 0 V
14 0 V
14 0 V
15 -1 V
14 0 V
15 0 V
14 0 V
1.000 UL
LT1
284 509 M
15 -1 V
14 -2 V
15 -2 V
14 -3 V
14 -3 V
15 -3 V
14 -4 V
15 -4 V
14 -3 V
14 -4 V
15 -3 V
14 -3 V
15 -2 V
14 -2 V
14 -1 V
15 -1 V
14 0 V
15 0 V
14 1 V
14 2 V
15 2 V
14 3 V
15 3 V
14 5 V
14 5 V
15 6 V
14 6 V
15 8 V
14 8 V
14 10 V
15 10 V
14 11 V
15 12 V
14 13 V
14 14 V
15 15 V
14 15 V
15 15 V
14 16 V
14 15 V
15 15 V
14 14 V
15 14 V
14 11 V
14 10 V
15 8 V
14 6 V
15 4 V
14 1 V
14 -1 V
15 -4 V
14 -6 V
15 -8 V
14 -10 V
14 -11 V
15 -14 V
14 -14 V
15 -15 V
14 -15 V
14 -16 V
15 -15 V
14 -15 V
15 -15 V
14 -14 V
14 -13 V
15 -12 V
14 -11 V
15 -10 V
14 -10 V
14 -8 V
15 -8 V
14 -6 V
15 -6 V
14 -5 V
14 -5 V
15 -3 V
14 -3 V
15 -2 V
14 -2 V
14 -1 V
15 0 V
14 0 V
15 1 V
14 1 V
14 2 V
15 2 V
14 3 V
15 3 V
14 4 V
14 3 V
15 4 V
14 4 V
15 3 V
14 3 V
14 3 V
15 2 V
14 2 V
15 1 V
14 1 V
1.000 UL
LT2
284 941 M
15 -22 V
14 -36 V
15 -47 V
14 -55 V
14 -61 V
15 -64 V
14 -63 V
15 -60 V
14 -55 V
14 -48 V
15 -41 V
14 -32 V
15 -24 V
14 -18 V
14 -10 V
15 -4 V
14 0 V
15 4 V
14 7 V
14 10 V
15 10 V
14 12 V
15 12 V
14 13 V
14 12 V
15 13 V
14 13 V
15 13 V
14 14 V
14 14 V
15 15 V
14 16 V
15 18 V
14 19 V
14 20 V
15 23 V
14 23 V
15 26 V
14 26 V
14 27 V
15 28 V
14 27 V
15 25 V
14 24 V
14 21 V
15 17 V
14 12 V
15 8 V
14 3 V
14 -3 V
15 -8 V
14 -12 V
15 -17 V
14 -21 V
14 -24 V
15 -25 V
14 -27 V
15 -28 V
14 -27 V
14 -26 V
15 -26 V
14 -23 V
15 -23 V
14 -20 V
14 -19 V
15 -18 V
14 -16 V
15 -15 V
14 -14 V
14 -14 V
15 -13 V
14 -13 V
15 -13 V
14 -12 V
14 -13 V
15 -12 V
14 -12 V
15 -10 V
14 -10 V
14 -7 V
15 -4 V
14 0 V
15 4 V
14 10 V
14 18 V
15 24 V
14 32 V
15 41 V
14 48 V
14 55 V
15 60 V
14 63 V
15 64 V
14 61 V
14 55 V
15 47 V
14 36 V
15 22 V
14 8 V
stroke
grestore
end
showpage
}}%
\put(990,30){\makebox(0,0){$\phi_--\phi^\prime_-$}}%
\put(990,-150){\makebox(0,0){\bf (b)}}%
\put(70,600){%
\special{ps: gsave currentpoint currentpoint translate
270 rotate neg exch neg exch translate}%
\makebox(0,0)[b]{\shortstack{${\cal P}^{(-)}(\phi_--\phi^\prime_-)$}}%
\special{ps: currentpoint grestore moveto}%
}%
\put(1710,120){\makebox(0,0){$\pi$}}%
\put(1350,120){\makebox(0,0){$\pi/2$}}%
\put(990,120){\makebox(0,0){$0$}}%
\put(630,120){\makebox(0,0){$-\pi/2$}}%
\put(270,120){\makebox(0,0){$-\pi$}}%
\put(240,1020){\makebox(0,0)[r]{1}}%
\put(240,740){\makebox(0,0)[r]{0.5}}%
\put(240,460){\makebox(0,0)[r]{0}}%
\put(240,180){\makebox(0,0)[r]{-0.5}}%
\end{picture}%
\endgroup
 

%% file: wigner_odd_state.tex
\begingroup%
  \makeatletter%
  \newcommand{\GNUPLOTspecial}{%
    \@sanitize\catcode`\%=14\relax\special}%
  \setlength{\unitlength}{0.1bp}%
\begin{picture}(3600,2160)(0,-500)%
\special{psfile=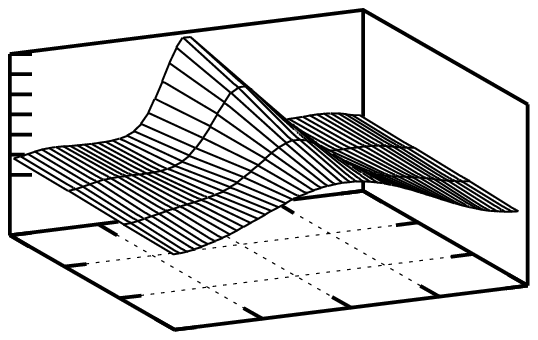 llx=0 lly=0 urx=720 ury=504 rwi=7200}
\put(1114,1594){\makebox(0,0){${\cal P}^{(-)}(\phi_--\phi^\prime_-)$}}%
\put(988,1463){\makebox(0,0)[r]{1}}%
\put(988,1405){\makebox(0,0)[r]{0.8}}%
\put(988,1347){\makebox(0,0)[r]{0.6}}%
\put(988,1289){\makebox(0,0)[r]{0.4}}%
\put(988,1231){\makebox(0,0)[r]{0.2}}%
\put(988,1173){\makebox(0,0)[r]{0}}%
\put(988,1115){\makebox(0,0)[r]{-0.2}}%
\put(2155,564){\makebox(0,0){$\phi_--\phi^\prime_-$}}%
\put(2155,384){\makebox(0,0){\bf (c)}}%
\put(1571,619){\makebox(0,0){$-\pi$}}%
\put(1825,650){\makebox(0,0){$-\pi/2$}}%
\put(2080,682){\makebox(0,0){0}}%
\put(2334,714){\makebox(0,0){$\pi/2$}}%
\put(2588,746){\makebox(0,0){$\pi$}}%
\put(1150,700){\makebox(0,0){$|\al|^2$}}%
\put(1004,928){\makebox(0,0)[r]{3}}%
\put(1162,838){\makebox(0,0)[r]{2}}%
\put(1320,747){\makebox(0,0)[r]{1}}%
\put(1478,656){\makebox(0,0)[r]{0}}%
\end{picture}%
\endgroup
 

%% file: p_odd_articol.tex
\begingroup%
  \makeatletter%
  \newcommand{\GNUPLOTspecial}{%
    \@sanitize\catcode`\%=14\relax\special}%
  \setlength{\unitlength}{0.1bp}%
{\GNUPLOTspecial{!
/gnudict 256 dict def
gnudict begin
/Color false def
/Solid false def
/gnulinewidth 5.000 def
/userlinewidth gnulinewidth def
/vshift -20 def
/dl {10 mul} def
/hpt_ 31.5 def
/vpt_ 31.5 def
/hpt hpt_ def
/vpt vpt_ def
/M {moveto} bind def
/L {lineto} bind def
/R {rmoveto} bind def
/V {rlineto} bind def
/vpt2 vpt 2 mul def
/hpt2 hpt 2 mul def
/Lshow { currentpoint stroke M
  0 vshift R show } def
/Rshow { currentpoint stroke M
  dup stringwidth pop neg vshift R show } def
/Cshow { currentpoint stroke M
  dup stringwidth pop -2 div vshift R show } def
/UP { dup vpt_ mul /vpt exch def hpt_ mul /hpt exch def
  /hpt2 hpt 2 mul def /vpt2 vpt 2 mul def } def
/DL { Color {setrgbcolor Solid {pop []} if 0 setdash }
 {pop pop pop Solid {pop []} if 0 setdash} ifelse } def
/BL { stroke userlinewidth 2 mul setlinewidth } def
/AL { stroke userlinewidth 2 div setlinewidth } def
/UL { dup gnulinewidth mul /userlinewidth exch def
      10 mul /udl exch def } def
/PL { stroke userlinewidth setlinewidth } def
/LTb { BL [] 0 0 0 DL } def
/LTa { AL [1 udl mul 2 udl mul] 0 setdash 0 0 0 setrgbcolor } def
/LT0 { PL [] 1 0 0 DL } def
/LT1 { PL [4 dl 2 dl] 0 1 0 DL } def
/LT2 { PL [2 dl 3 dl] 0 0 1 DL } def
/LT3 { PL [1 dl 1.5 dl] 1 0 1 DL } def
/LT4 { PL [5 dl 2 dl 1 dl 2 dl] 0 1 1 DL } def
/LT5 { PL [4 dl 3 dl 1 dl 3 dl] 1 1 0 DL } def
/LT6 { PL [2 dl 2 dl 2 dl 4 dl] 0 0 0 DL } def
/LT7 { PL [2 dl 2 dl 2 dl 2 dl 2 dl 4 dl] 1 0.3 0 DL } def
/LT8 { PL [2 dl 2 dl 2 dl 2 dl 2 dl 2 dl 2 dl 4 dl] 0.5 0.5 0.5 DL } def
/Pnt { stroke [] 0 setdash
   gsave 1 setlinecap M 0 0 V stroke grestore } def
/Dia { stroke [] 0 setdash 2 copy vpt add M
  hpt neg vpt neg V hpt vpt neg V
  hpt vpt V hpt neg vpt V closepath stroke
  Pnt } def
/Pls { stroke [] 0 setdash vpt sub M 0 vpt2 V
  currentpoint stroke M
  hpt neg vpt neg R hpt2 0 V stroke
  } def
/Box { stroke [] 0 setdash 2 copy exch hpt sub exch vpt add M
  0 vpt2 neg V hpt2 0 V 0 vpt2 V
  hpt2 neg 0 V closepath stroke
  Pnt } def
/Crs { stroke [] 0 setdash exch hpt sub exch vpt add M
  hpt2 vpt2 neg V currentpoint stroke M
  hpt2 neg 0 R hpt2 vpt2 V stroke } def
/TriU { stroke [] 0 setdash 2 copy vpt 1.12 mul add M
  hpt neg vpt -1.62 mul V
  hpt 2 mul 0 V
  hpt neg vpt 1.62 mul V closepath stroke
  Pnt  } def
/Star { 2 copy Pls Crs } def
/BoxF { stroke [] 0 setdash exch hpt sub exch vpt add M
  0 vpt2 neg V  hpt2 0 V  0 vpt2 V
  hpt2 neg 0 V  closepath fill } def
/TriUF { stroke [] 0 setdash vpt 1.12 mul add M
  hpt neg vpt -1.62 mul V
  hpt 2 mul 0 V
  hpt neg vpt 1.62 mul V closepath fill } def
/TriD { stroke [] 0 setdash 2 copy vpt 1.12 mul sub M
  hpt neg vpt 1.62 mul V
  hpt 2 mul 0 V
  hpt neg vpt -1.62 mul V closepath stroke
  Pnt  } def
/TriDF { stroke [] 0 setdash vpt 1.12 mul sub M
  hpt neg vpt 1.62 mul V
  hpt 2 mul 0 V
  hpt neg vpt -1.62 mul V closepath fill} def
/DiaF { stroke [] 0 setdash vpt add M
  hpt neg vpt neg V hpt vpt neg V
  hpt vpt V hpt neg vpt V closepath fill } def
/Pent { stroke [] 0 setdash 2 copy gsave
  translate 0 hpt M 4 {72 rotate 0 hpt L} repeat
  closepath stroke grestore Pnt } def
/PentF { stroke [] 0 setdash gsave
  translate 0 hpt M 4 {72 rotate 0 hpt L} repeat
  closepath fill grestore } def
/Circle { stroke [] 0 setdash 2 copy
  hpt 0 360 arc stroke Pnt } def
/CircleF { stroke [] 0 setdash hpt 0 360 arc fill } def
/C0 { BL [] 0 setdash 2 copy moveto vpt 90 450  arc } bind def
/C1 { BL [] 0 setdash 2 copy        moveto
       2 copy  vpt 0 90 arc closepath fill
               vpt 0 360 arc closepath } bind def
/C2 { BL [] 0 setdash 2 copy moveto
       2 copy  vpt 90 180 arc closepath fill
               vpt 0 360 arc closepath } bind def
/C3 { BL [] 0 setdash 2 copy moveto
       2 copy  vpt 0 180 arc closepath fill
               vpt 0 360 arc closepath } bind def
/C4 { BL [] 0 setdash 2 copy moveto
       2 copy  vpt 180 270 arc closepath fill
               vpt 0 360 arc closepath } bind def
/C5 { BL [] 0 setdash 2 copy moveto
       2 copy  vpt 0 90 arc
       2 copy moveto
       2 copy  vpt 180 270 arc closepath fill
               vpt 0 360 arc } bind def
/C6 { BL [] 0 setdash 2 copy moveto
      2 copy  vpt 90 270 arc closepath fill
              vpt 0 360 arc closepath } bind def
/C7 { BL [] 0 setdash 2 copy moveto
      2 copy  vpt 0 270 arc closepath fill
              vpt 0 360 arc closepath } bind def
/C8 { BL [] 0 setdash 2 copy moveto
      2 copy vpt 270 360 arc closepath fill
              vpt 0 360 arc closepath } bind def
/C9 { BL [] 0 setdash 2 copy moveto
      2 copy  vpt 270 450 arc closepath fill
              vpt 0 360 arc closepath } bind def
/C10 { BL [] 0 setdash 2 copy 2 copy moveto vpt 270 360 arc closepath fill
       2 copy moveto
       2 copy vpt 90 180 arc closepath fill
               vpt 0 360 arc closepath } bind def
/C11 { BL [] 0 setdash 2 copy moveto
       2 copy  vpt 0 180 arc closepath fill
       2 copy moveto
       2 copy  vpt 270 360 arc closepath fill
               vpt 0 360 arc closepath } bind def
/C12 { BL [] 0 setdash 2 copy moveto
       2 copy  vpt 180 360 arc closepath fill
               vpt 0 360 arc closepath } bind def
/C13 { BL [] 0 setdash  2 copy moveto
       2 copy  vpt 0 90 arc closepath fill
       2 copy moveto
       2 copy  vpt 180 360 arc closepath fill
               vpt 0 360 arc closepath } bind def
/C14 { BL [] 0 setdash 2 copy moveto
       2 copy  vpt 90 360 arc closepath fill
               vpt 0 360 arc } bind def
/C15 { BL [] 0 setdash 2 copy vpt 0 360 arc closepath fill
               vpt 0 360 arc closepath } bind def
/Rec   { newpath 4 2 roll moveto 1 index 0 rlineto 0 exch rlineto
       neg 0 rlineto closepath } bind def
/Square { dup Rec } bind def
/Bsquare { vpt sub exch vpt sub exch vpt2 Square } bind def
/S0 { BL [] 0 setdash 2 copy moveto 0 vpt rlineto BL Bsquare } bind def
/S1 { BL [] 0 setdash 2 copy vpt Square fill Bsquare } bind def
/S2 { BL [] 0 setdash 2 copy exch vpt sub exch vpt Square fill Bsquare } bind def
/S3 { BL [] 0 setdash 2 copy exch vpt sub exch vpt2 vpt Rec fill Bsquare } bind def
/S4 { BL [] 0 setdash 2 copy exch vpt sub exch vpt sub vpt Square fill Bsquare } bind def
/S5 { BL [] 0 setdash 2 copy 2 copy vpt Square fill
       exch vpt sub exch vpt sub vpt Square fill Bsquare } bind def
/S6 { BL [] 0 setdash 2 copy exch vpt sub exch vpt sub vpt vpt2 Rec fill Bsquare } bind def
/S7 { BL [] 0 setdash 2 copy exch vpt sub exch vpt sub vpt vpt2 Rec fill
       2 copy vpt Square fill
       Bsquare } bind def
/S8 { BL [] 0 setdash 2 copy vpt sub vpt Square fill Bsquare } bind def
/S9 { BL [] 0 setdash 2 copy vpt sub vpt vpt2 Rec fill Bsquare } bind def
/S10 { BL [] 0 setdash 2 copy vpt sub vpt Square fill 2 copy exch vpt sub exch vpt Square fill
       Bsquare } bind def
/S11 { BL [] 0 setdash 2 copy vpt sub vpt Square fill 2 copy exch vpt sub exch vpt2 vpt Rec fill
       Bsquare } bind def
/S12 { BL [] 0 setdash 2 copy exch vpt sub exch vpt sub vpt2 vpt Rec fill Bsquare } bind def
/S13 { BL [] 0 setdash 2 copy exch vpt sub exch vpt sub vpt2 vpt Rec fill
       2 copy vpt Square fill Bsquare } bind def
/S14 { BL [] 0 setdash 2 copy exch vpt sub exch vpt sub vpt2 vpt Rec fill
       2 copy exch vpt sub exch vpt Square fill Bsquare } bind def
/S15 { BL [] 0 setdash 2 copy Bsquare fill Bsquare } bind def
/D0 { gsave translate 45 rotate 0 0 S0 stroke grestore } bind def
/D1 { gsave translate 45 rotate 0 0 S1 stroke grestore } bind def
/D2 { gsave translate 45 rotate 0 0 S2 stroke grestore } bind def
/D3 { gsave translate 45 rotate 0 0 S3 stroke grestore } bind def
/D4 { gsave translate 45 rotate 0 0 S4 stroke grestore } bind def
/D5 { gsave translate 45 rotate 0 0 S5 stroke grestore } bind def
/D6 { gsave translate 45 rotate 0 0 S6 stroke grestore } bind def
/D7 { gsave translate 45 rotate 0 0 S7 stroke grestore } bind def
/D8 { gsave translate 45 rotate 0 0 S8 stroke grestore } bind def
/D9 { gsave translate 45 rotate 0 0 S9 stroke grestore } bind def
/D10 { gsave translate 45 rotate 0 0 S10 stroke grestore } bind def
/D11 { gsave translate 45 rotate 0 0 S11 stroke grestore } bind def
/D12 { gsave translate 45 rotate 0 0 S12 stroke grestore } bind def
/D13 { gsave translate 45 rotate 0 0 S13 stroke grestore } bind def
/D14 { gsave translate 45 rotate 0 0 S14 stroke grestore } bind def
/D15 { gsave translate 45 rotate 0 0 S15 stroke grestore } bind def
/DiaE { stroke [] 0 setdash vpt add M
  hpt neg vpt neg V hpt vpt neg V
  hpt vpt V hpt neg vpt V closepath stroke } def
/BoxE { stroke [] 0 setdash exch hpt sub exch vpt add M
  0 vpt2 neg V hpt2 0 V 0 vpt2 V
  hpt2 neg 0 V closepath stroke } def
/TriUE { stroke [] 0 setdash vpt 1.12 mul add M
  hpt neg vpt -1.62 mul V
  hpt 2 mul 0 V
  hpt neg vpt 1.62 mul V closepath stroke } def
/TriDE { stroke [] 0 setdash vpt 1.12 mul sub M
  hpt neg vpt 1.62 mul V
  hpt 2 mul 0 V
  hpt neg vpt -1.62 mul V closepath stroke } def
/PentE { stroke [] 0 setdash gsave
  translate 0 hpt M 4 {72 rotate 0 hpt L} repeat
  closepath stroke grestore } def
/CircE { stroke [] 0 setdash 
  hpt 0 360 arc stroke } def
/Opaque { gsave closepath 1 setgray fill grestore 0 setgray closepath } def
/DiaW { stroke [] 0 setdash vpt add M
  hpt neg vpt neg V hpt vpt neg V
  hpt vpt V hpt neg vpt V Opaque stroke } def
/BoxW { stroke [] 0 setdash exch hpt sub exch vpt add M
  0 vpt2 neg V hpt2 0 V 0 vpt2 V
  hpt2 neg 0 V Opaque stroke } def
/TriUW { stroke [] 0 setdash vpt 1.12 mul add M
  hpt neg vpt -1.62 mul V
  hpt 2 mul 0 V
  hpt neg vpt 1.62 mul V Opaque stroke } def
/TriDW { stroke [] 0 setdash vpt 1.12 mul sub M
  hpt neg vpt 1.62 mul V
  hpt 2 mul 0 V
  hpt neg vpt -1.62 mul V Opaque stroke } def
/PentW { stroke [] 0 setdash gsave
  translate 0 hpt M 4 {72 rotate 0 hpt L} repeat
  Opaque stroke grestore } def
/CircW { stroke [] 0 setdash 
  hpt 0 360 arc Opaque stroke } def
/BoxFill { gsave Rec 1 setgray fill grestore } def
/Symbol-Oblique /Symbol findfont [1 0 .167 1 0 0] makefont
dup length dict begin {1 index /FID eq {pop pop} {def} ifelse} forall
currentdict end definefont
end
}}%
\begin{picture}(1800,1080)(0,-1040)%
{\GNUPLOTspecial{"
gnudict begin
gsave
0 0 translate
0.100 0.100 scale
0 setgray
newpath
1.000 UL
LTb
1.000 UL
LTa
270 180 M
1440 0 V
1.000 UL
LTb
270 180 M
63 0 V
1377 0 R
-63 0 V
1.000 UL
LTa
270 390 M
1440 0 V
1.000 UL
LTb
270 390 M
63 0 V
1377 0 R
-63 0 V
1.000 UL
LTa
270 600 M
1440 0 V
1.000 UL
LTb
270 600 M
63 0 V
1377 0 R
-63 0 V
1.000 UL
LTa
270 810 M
1440 0 V
1.000 UL
LTb
270 810 M
63 0 V
1377 0 R
-63 0 V
1.000 UL
LTa
270 1020 M
1440 0 V
1.000 UL
LTb
270 1020 M
63 0 V
1377 0 R
-63 0 V
1.000 UL
LTa
270 180 M
0 840 V
1.000 UL
LTb
270 180 M
0 63 V
0 777 R
0 -63 V
1.000 UL
LTa
630 180 M
0 840 V
1.000 UL
LTb
630 180 M
0 63 V
0 777 R
0 -63 V
1.000 UL
LTa
990 180 M
0 840 V
1.000 UL
LTb
990 180 M
0 63 V
0 777 R
0 -63 V
1.000 UL
LTa
1350 180 M
0 840 V
1.000 UL
LTb
1350 180 M
0 63 V
0 777 R
0 -63 V
1.000 UL
LTa
1710 180 M
0 840 V
1.000 UL
LTb
1710 180 M
0 63 V
0 777 R
0 -63 V
1.000 UL
LTb
270 180 M
1440 0 V
0 840 V
-1440 0 V
0 -840 V
1.000 UL
LT0
284 609 M
15 0 V
14 0 V
15 1 V
14 0 V
14 1 V
15 1 V
14 0 V
15 1 V
14 2 V
14 1 V
15 1 V
14 2 V
15 1 V
14 2 V
14 2 V
15 2 V
14 3 V
15 2 V
14 3 V
14 2 V
15 4 V
14 3 V
15 3 V
14 4 V
14 4 V
15 4 V
14 5 V
15 5 V
14 5 V
14 5 V
15 5 V
14 6 V
15 5 V
14 6 V
14 6 V
15 6 V
14 6 V
15 6 V
14 5 V
14 5 V
15 5 V
14 5 V
15 4 V
14 4 V
14 3 V
15 2 V
14 2 V
15 1 V
14 0 V
14 0 V
15 -1 V
14 -2 V
15 -2 V
14 -3 V
14 -4 V
15 -4 V
14 -5 V
15 -5 V
14 -5 V
14 -5 V
15 -6 V
14 -6 V
15 -6 V
14 -6 V
14 -6 V
15 -5 V
14 -6 V
15 -5 V
14 -5 V
14 -5 V
15 -5 V
14 -5 V
15 -4 V
14 -4 V
14 -4 V
15 -3 V
14 -3 V
15 -4 V
14 -2 V
14 -3 V
15 -2 V
14 -3 V
15 -2 V
14 -2 V
14 -2 V
15 -1 V
14 -2 V
15 -1 V
14 -1 V
14 -2 V
15 -1 V
14 0 V
15 -1 V
14 -1 V
14 0 V
15 -1 V
14 0 V
15 0 V
14 0 V
1.000 UL
LT1
284 569 M
15 1 V
14 1 V
15 2 V
14 3 V
14 3 V
15 3 V
14 3 V
15 4 V
14 3 V
14 4 V
15 3 V
14 3 V
15 4 V
14 3 V
14 2 V
15 3 V
14 3 V
15 2 V
14 3 V
14 3 V
15 2 V
14 3 V
15 3 V
14 4 V
14 4 V
15 4 V
14 5 V
15 5 V
14 6 V
14 7 V
15 8 V
14 8 V
15 9 V
14 9 V
14 11 V
15 10 V
14 12 V
15 11 V
14 12 V
14 12 V
15 11 V
14 11 V
15 10 V
14 9 V
14 7 V
15 6 V
14 5 V
15 3 V
14 1 V
14 -1 V
15 -3 V
14 -5 V
15 -6 V
14 -7 V
14 -9 V
15 -10 V
14 -11 V
15 -11 V
14 -12 V
14 -12 V
15 -11 V
14 -12 V
15 -10 V
14 -11 V
14 -9 V
15 -9 V
14 -8 V
15 -8 V
14 -7 V
14 -6 V
15 -5 V
14 -5 V
15 -4 V
14 -4 V
14 -4 V
15 -3 V
14 -3 V
15 -2 V
14 -3 V
14 -3 V
15 -2 V
14 -3 V
15 -3 V
14 -2 V
14 -3 V
15 -4 V
14 -3 V
15 -3 V
14 -4 V
14 -3 V
15 -4 V
14 -3 V
15 -3 V
14 -3 V
14 -3 V
15 -2 V
14 -1 V
15 -1 V
14 0 V
1.000 UL
LT2
284 227 M
15 17 V
14 28 V
15 37 V
14 43 V
14 47 V
15 50 V
14 49 V
15 47 V
14 43 V
14 38 V
15 31 V
14 26 V
15 19 V
14 14 V
14 9 V
15 4 V
14 0 V
15 -2 V
14 -4 V
14 -6 V
15 -6 V
14 -7 V
15 -6 V
14 -6 V
14 -5 V
15 -4 V
14 -3 V
15 -2 V
14 0 V
14 2 V
15 3 V
14 5 V
15 7 V
14 9 V
14 11 V
15 13 V
14 15 V
15 16 V
14 18 V
14 19 V
15 19 V
14 19 V
15 19 V
14 17 V
14 15 V
15 13 V
14 9 V
15 6 V
14 2 V
14 -2 V
15 -6 V
14 -9 V
15 -13 V
14 -15 V
14 -17 V
15 -19 V
14 -19 V
15 -19 V
14 -19 V
14 -18 V
15 -16 V
14 -15 V
15 -13 V
14 -11 V
14 -9 V
15 -7 V
14 -5 V
15 -3 V
14 -2 V
14 0 V
15 2 V
14 3 V
15 4 V
14 5 V
14 6 V
15 6 V
14 7 V
15 6 V
14 6 V
14 4 V
15 2 V
14 0 V
15 -4 V
14 -9 V
14 -14 V
15 -19 V
14 -26 V
15 -31 V
14 -38 V
14 -43 V
15 -47 V
14 -49 V
15 -50 V
14 -47 V
14 -43 V
15 -37 V
14 -28 V
15 -17 V
14 -6 V
stroke
grestore
end
showpage
}}%
\put(990,30){\makebox(0,0){$\phi_--\phi^\prime_-$}}%
\put(990,-150){\makebox(0,0){\bf (d)}}%
\put(70,600){%
\special{ps: gsave currentpoint currentpoint translate
270 rotate neg exch neg exch translate}%
\makebox(0,0)[b]{\shortstack{${\cal P}^{(-)}(\phi_--\phi^\prime_-)$}}%
\special{ps: currentpoint grestore moveto}%
}%
\put(1710,120){\makebox(0,0){$\pi$}}%
\put(1350,120){\makebox(0,0){$\pi/2$}}%
\put(990,120){\makebox(0,0){$0$}}%
\put(630,120){\makebox(0,0){$-\pi/2$}}%
\put(270,120){\makebox(0,0){$-\pi$}}%
\put(240,1020){\makebox(0,0)[r]{1}}%
\put(240,810){\makebox(0,0)[r]{0.5}}%
\put(240,600){\makebox(0,0)[r]{0}}%
\put(240,390){\makebox(0,0)[r]{-0.5}}%
\put(240,180){\makebox(0,0)[r]{-1}}%
\end{picture}%
\endgroup
 

%% file: pw_sum_even.tex
\begingroup%
  \makeatletter%
  \newcommand{\GNUPLOTspecial}{%
    \@sanitize\catcode`\%=14\relax\special}%
  \setlength{\unitlength}{0.1bp}%
\begin{picture}(3600,2160)(0,-500)%
\special{psfile=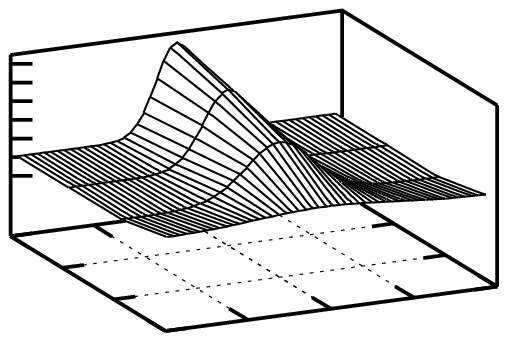 llx=0 lly=0 urx=720 ury=504 rwi=7200}
\put(1114,1594){\makebox(0,0){${\cal P}^{(+)}(\phi_+-\phi^\prime_+)$}}%
\put(1033,1417){\makebox(0,0)[r]{1}}%
\put(1033,1363){\makebox(0,0)[r]{0.8}}%
\put(1033,1310){\makebox(0,0)[r]{0.6}}%
\put(1033,1256){\makebox(0,0)[r]{0.4}}%
\put(1033,1202){\makebox(0,0)[r]{0.2}}%
\put(1033,1149){\makebox(0,0)[r]{0}}%
\put(1033,1095){\makebox(0,0)[r]{-0.2}}%
\put(2134,560){\makebox(0,0){$\phi_+-\phi^\prime_+$}}%
\put(2134,404){\makebox(0,0){\bf (a)}}%
\put(1570,596){\makebox(0,0){$-\pi$}}%
\put(1808,628){\makebox(0,0){$-\pi/2$}}%
\put(2047,660){\makebox(0,0){0}}%
\put(2286,692){\makebox(0,0){$\pi/2$}}%
\put(2525,724){\makebox(0,0){$\pi$}}%
\put(1196,674){\makebox(0,0){$|\al|^2$}}%
\put(1069,903){\makebox(0,0)[r]{3}}%
\put(1218,812){\makebox(0,0)[r]{2}}%
\put(1366,721){\makebox(0,0)[r]{1}}%
\put(1515,630){\makebox(0,0)[r]{0}}%
\end{picture}%
\endgroup
 

%% file: pp_sum_even.tex
\begingroup%
  \makeatletter%
  \newcommand{\GNUPLOTspecial}{%
    \@sanitize\catcode`\%=14\relax\special}%
  \setlength{\unitlength}{0.1bp}%
{\GNUPLOTspecial{!
/gnudict 256 dict def
gnudict begin
/Color false def
/Solid false def
/gnulinewidth 5.000 def
/userlinewidth gnulinewidth def
/vshift -20 def
/dl {10 mul} def
/hpt_ 31.5 def
/vpt_ 31.5 def
/hpt hpt_ def
/vpt vpt_ def
/M {moveto} bind def
/L {lineto} bind def
/R {rmoveto} bind def
/V {rlineto} bind def
/vpt2 vpt 2 mul def
/hpt2 hpt 2 mul def
/Lshow { currentpoint stroke M
  0 vshift R show } def
/Rshow { currentpoint stroke M
  dup stringwidth pop neg vshift R show } def
/Cshow { currentpoint stroke M
  dup stringwidth pop -2 div vshift R show } def
/UP { dup vpt_ mul /vpt exch def hpt_ mul /hpt exch def
  /hpt2 hpt 2 mul def /vpt2 vpt 2 mul def } def
/DL { Color {setrgbcolor Solid {pop []} if 0 setdash }
 {pop pop pop Solid {pop []} if 0 setdash} ifelse } def
/BL { stroke userlinewidth 2 mul setlinewidth } def
/AL { stroke userlinewidth 2 div setlinewidth } def
/UL { dup gnulinewidth mul /userlinewidth exch def
      10 mul /udl exch def } def
/PL { stroke userlinewidth setlinewidth } def
/LTb { BL [] 0 0 0 DL } def
/LTa { AL [1 udl mul 2 udl mul] 0 setdash 0 0 0 setrgbcolor } def
/LT0 { PL [] 1 0 0 DL } def
/LT1 { PL [4 dl 2 dl] 0 1 0 DL } def
/LT2 { PL [2 dl 3 dl] 0 0 1 DL } def
/LT3 { PL [1 dl 1.5 dl] 1 0 1 DL } def
/LT4 { PL [5 dl 2 dl 1 dl 2 dl] 0 1 1 DL } def
/LT5 { PL [4 dl 3 dl 1 dl 3 dl] 1 1 0 DL } def
/LT6 { PL [2 dl 2 dl 2 dl 4 dl] 0 0 0 DL } def
/LT7 { PL [2 dl 2 dl 2 dl 2 dl 2 dl 4 dl] 1 0.3 0 DL } def
/LT8 { PL [2 dl 2 dl 2 dl 2 dl 2 dl 2 dl 2 dl 4 dl] 0.5 0.5 0.5 DL } def
/Pnt { stroke [] 0 setdash
   gsave 1 setlinecap M 0 0 V stroke grestore } def
/Dia { stroke [] 0 setdash 2 copy vpt add M
  hpt neg vpt neg V hpt vpt neg V
  hpt vpt V hpt neg vpt V closepath stroke
  Pnt } def
/Pls { stroke [] 0 setdash vpt sub M 0 vpt2 V
  currentpoint stroke M
  hpt neg vpt neg R hpt2 0 V stroke
  } def
/Box { stroke [] 0 setdash 2 copy exch hpt sub exch vpt add M
  0 vpt2 neg V hpt2 0 V 0 vpt2 V
  hpt2 neg 0 V closepath stroke
  Pnt } def
/Crs { stroke [] 0 setdash exch hpt sub exch vpt add M
  hpt2 vpt2 neg V currentpoint stroke M
  hpt2 neg 0 R hpt2 vpt2 V stroke } def
/TriU { stroke [] 0 setdash 2 copy vpt 1.12 mul add M
  hpt neg vpt -1.62 mul V
  hpt 2 mul 0 V
  hpt neg vpt 1.62 mul V closepath stroke
  Pnt  } def
/Star { 2 copy Pls Crs } def
/BoxF { stroke [] 0 setdash exch hpt sub exch vpt add M
  0 vpt2 neg V  hpt2 0 V  0 vpt2 V
  hpt2 neg 0 V  closepath fill } def
/TriUF { stroke [] 0 setdash vpt 1.12 mul add M
  hpt neg vpt -1.62 mul V
  hpt 2 mul 0 V
  hpt neg vpt 1.62 mul V closepath fill } def
/TriD { stroke [] 0 setdash 2 copy vpt 1.12 mul sub M
  hpt neg vpt 1.62 mul V
  hpt 2 mul 0 V
  hpt neg vpt -1.62 mul V closepath stroke
  Pnt  } def
/TriDF { stroke [] 0 setdash vpt 1.12 mul sub M
  hpt neg vpt 1.62 mul V
  hpt 2 mul 0 V
  hpt neg vpt -1.62 mul V closepath fill} def
/DiaF { stroke [] 0 setdash vpt add M
  hpt neg vpt neg V hpt vpt neg V
  hpt vpt V hpt neg vpt V closepath fill } def
/Pent { stroke [] 0 setdash 2 copy gsave
  translate 0 hpt M 4 {72 rotate 0 hpt L} repeat
  closepath stroke grestore Pnt } def
/PentF { stroke [] 0 setdash gsave
  translate 0 hpt M 4 {72 rotate 0 hpt L} repeat
  closepath fill grestore } def
/Circle { stroke [] 0 setdash 2 copy
  hpt 0 360 arc stroke Pnt } def
/CircleF { stroke [] 0 setdash hpt 0 360 arc fill } def
/C0 { BL [] 0 setdash 2 copy moveto vpt 90 450  arc } bind def
/C1 { BL [] 0 setdash 2 copy        moveto
       2 copy  vpt 0 90 arc closepath fill
               vpt 0 360 arc closepath } bind def
/C2 { BL [] 0 setdash 2 copy moveto
       2 copy  vpt 90 180 arc closepath fill
               vpt 0 360 arc closepath } bind def
/C3 { BL [] 0 setdash 2 copy moveto
       2 copy  vpt 0 180 arc closepath fill
               vpt 0 360 arc closepath } bind def
/C4 { BL [] 0 setdash 2 copy moveto
       2 copy  vpt 180 270 arc closepath fill
               vpt 0 360 arc closepath } bind def
/C5 { BL [] 0 setdash 2 copy moveto
       2 copy  vpt 0 90 arc
       2 copy moveto
       2 copy  vpt 180 270 arc closepath fill
               vpt 0 360 arc } bind def
/C6 { BL [] 0 setdash 2 copy moveto
      2 copy  vpt 90 270 arc closepath fill
              vpt 0 360 arc closepath } bind def
/C7 { BL [] 0 setdash 2 copy moveto
      2 copy  vpt 0 270 arc closepath fill
              vpt 0 360 arc closepath } bind def
/C8 { BL [] 0 setdash 2 copy moveto
      2 copy vpt 270 360 arc closepath fill
              vpt 0 360 arc closepath } bind def
/C9 { BL [] 0 setdash 2 copy moveto
      2 copy  vpt 270 450 arc closepath fill
              vpt 0 360 arc closepath } bind def
/C10 { BL [] 0 setdash 2 copy 2 copy moveto vpt 270 360 arc closepath fill
       2 copy moveto
       2 copy vpt 90 180 arc closepath fill
               vpt 0 360 arc closepath } bind def
/C11 { BL [] 0 setdash 2 copy moveto
       2 copy  vpt 0 180 arc closepath fill
       2 copy moveto
       2 copy  vpt 270 360 arc closepath fill
               vpt 0 360 arc closepath } bind def
/C12 { BL [] 0 setdash 2 copy moveto
       2 copy  vpt 180 360 arc closepath fill
               vpt 0 360 arc closepath } bind def
/C13 { BL [] 0 setdash  2 copy moveto
       2 copy  vpt 0 90 arc closepath fill
       2 copy moveto
       2 copy  vpt 180 360 arc closepath fill
               vpt 0 360 arc closepath } bind def
/C14 { BL [] 0 setdash 2 copy moveto
       2 copy  vpt 90 360 arc closepath fill
               vpt 0 360 arc } bind def
/C15 { BL [] 0 setdash 2 copy vpt 0 360 arc closepath fill
               vpt 0 360 arc closepath } bind def
/Rec   { newpath 4 2 roll moveto 1 index 0 rlineto 0 exch rlineto
       neg 0 rlineto closepath } bind def
/Square { dup Rec } bind def
/Bsquare { vpt sub exch vpt sub exch vpt2 Square } bind def
/S0 { BL [] 0 setdash 2 copy moveto 0 vpt rlineto BL Bsquare } bind def
/S1 { BL [] 0 setdash 2 copy vpt Square fill Bsquare } bind def
/S2 { BL [] 0 setdash 2 copy exch vpt sub exch vpt Square fill Bsquare } bind def
/S3 { BL [] 0 setdash 2 copy exch vpt sub exch vpt2 vpt Rec fill Bsquare } bind def
/S4 { BL [] 0 setdash 2 copy exch vpt sub exch vpt sub vpt Square fill Bsquare } bind def
/S5 { BL [] 0 setdash 2 copy 2 copy vpt Square fill
       exch vpt sub exch vpt sub vpt Square fill Bsquare } bind def
/S6 { BL [] 0 setdash 2 copy exch vpt sub exch vpt sub vpt vpt2 Rec fill Bsquare } bind def
/S7 { BL [] 0 setdash 2 copy exch vpt sub exch vpt sub vpt vpt2 Rec fill
       2 copy vpt Square fill
       Bsquare } bind def
/S8 { BL [] 0 setdash 2 copy vpt sub vpt Square fill Bsquare } bind def
/S9 { BL [] 0 setdash 2 copy vpt sub vpt vpt2 Rec fill Bsquare } bind def
/S10 { BL [] 0 setdash 2 copy vpt sub vpt Square fill 2 copy exch vpt sub exch vpt Square fill
       Bsquare } bind def
/S11 { BL [] 0 setdash 2 copy vpt sub vpt Square fill 2 copy exch vpt sub exch vpt2 vpt Rec fill
       Bsquare } bind def
/S12 { BL [] 0 setdash 2 copy exch vpt sub exch vpt sub vpt2 vpt Rec fill Bsquare } bind def
/S13 { BL [] 0 setdash 2 copy exch vpt sub exch vpt sub vpt2 vpt Rec fill
       2 copy vpt Square fill Bsquare } bind def
/S14 { BL [] 0 setdash 2 copy exch vpt sub exch vpt sub vpt2 vpt Rec fill
       2 copy exch vpt sub exch vpt Square fill Bsquare } bind def
/S15 { BL [] 0 setdash 2 copy Bsquare fill Bsquare } bind def
/D0 { gsave translate 45 rotate 0 0 S0 stroke grestore } bind def
/D1 { gsave translate 45 rotate 0 0 S1 stroke grestore } bind def
/D2 { gsave translate 45 rotate 0 0 S2 stroke grestore } bind def
/D3 { gsave translate 45 rotate 0 0 S3 stroke grestore } bind def
/D4 { gsave translate 45 rotate 0 0 S4 stroke grestore } bind def
/D5 { gsave translate 45 rotate 0 0 S5 stroke grestore } bind def
/D6 { gsave translate 45 rotate 0 0 S6 stroke grestore } bind def
/D7 { gsave translate 45 rotate 0 0 S7 stroke grestore } bind def
/D8 { gsave translate 45 rotate 0 0 S8 stroke grestore } bind def
/D9 { gsave translate 45 rotate 0 0 S9 stroke grestore } bind def
/D10 { gsave translate 45 rotate 0 0 S10 stroke grestore } bind def
/D11 { gsave translate 45 rotate 0 0 S11 stroke grestore } bind def
/D12 { gsave translate 45 rotate 0 0 S12 stroke grestore } bind def
/D13 { gsave translate 45 rotate 0 0 S13 stroke grestore } bind def
/D14 { gsave translate 45 rotate 0 0 S14 stroke grestore } bind def
/D15 { gsave translate 45 rotate 0 0 S15 stroke grestore } bind def
/DiaE { stroke [] 0 setdash vpt add M
  hpt neg vpt neg V hpt vpt neg V
  hpt vpt V hpt neg vpt V closepath stroke } def
/BoxE { stroke [] 0 setdash exch hpt sub exch vpt add M
  0 vpt2 neg V hpt2 0 V 0 vpt2 V
  hpt2 neg 0 V closepath stroke } def
/TriUE { stroke [] 0 setdash vpt 1.12 mul add M
  hpt neg vpt -1.62 mul V
  hpt 2 mul 0 V
  hpt neg vpt 1.62 mul V closepath stroke } def
/TriDE { stroke [] 0 setdash vpt 1.12 mul sub M
  hpt neg vpt 1.62 mul V
  hpt 2 mul 0 V
  hpt neg vpt -1.62 mul V closepath stroke } def
/PentE { stroke [] 0 setdash gsave
  translate 0 hpt M 4 {72 rotate 0 hpt L} repeat
  closepath stroke grestore } def
/CircE { stroke [] 0 setdash 
  hpt 0 360 arc stroke } def
/Opaque { gsave closepath 1 setgray fill grestore 0 setgray closepath } def
/DiaW { stroke [] 0 setdash vpt add M
  hpt neg vpt neg V hpt vpt neg V
  hpt vpt V hpt neg vpt V Opaque stroke } def
/BoxW { stroke [] 0 setdash exch hpt sub exch vpt add M
  0 vpt2 neg V hpt2 0 V 0 vpt2 V
  hpt2 neg 0 V Opaque stroke } def
/TriUW { stroke [] 0 setdash vpt 1.12 mul add M
  hpt neg vpt -1.62 mul V
  hpt 2 mul 0 V
  hpt neg vpt 1.62 mul V Opaque stroke } def
/TriDW { stroke [] 0 setdash vpt 1.12 mul sub M
  hpt neg vpt 1.62 mul V
  hpt 2 mul 0 V
  hpt neg vpt -1.62 mul V Opaque stroke } def
/PentW { stroke [] 0 setdash gsave
  translate 0 hpt M 4 {72 rotate 0 hpt L} repeat
  Opaque stroke grestore } def
/CircW { stroke [] 0 setdash 
  hpt 0 360 arc Opaque stroke } def
/BoxFill { gsave Rec 1 setgray fill grestore } def
/Symbol-Oblique /Symbol findfont [1 0 .167 1 0 0] makefont
dup length dict begin {1 index /FID eq {pop pop} {def} ifelse} forall
currentdict end definefont
end
}}%
\begin{picture}(1800,1080)(0,-1040)%
{\GNUPLOTspecial{"
gnudict begin
gsave
0 0 translate
0.100 0.100 scale
0 setgray
newpath
1.000 UL
LTb
1.000 UL
LTa
300 180 M
1410 0 V
1.000 UL
LTb
300 180 M
63 0 V
1347 0 R
-63 0 V
1.000 UL
LTa
300 336 M
1410 0 V
1.000 UL
LTb
300 336 M
63 0 V
1347 0 R
-63 0 V
1.000 UL
LTa
300 491 M
1410 0 V
1.000 UL
LTb
300 491 M
63 0 V
1347 0 R
-63 0 V
1.000 UL
LTa
300 647 M
1410 0 V
1.000 UL
LTb
300 647 M
63 0 V
1347 0 R
-63 0 V
1.000 UL
LTa
300 802 M
1410 0 V
1.000 UL
LTb
300 802 M
63 0 V
1347 0 R
-63 0 V
1.000 UL
LTa
300 958 M
1410 0 V
1.000 UL
LTb
300 958 M
63 0 V
1347 0 R
-63 0 V
1.000 UL
LTa
300 180 M
0 840 V
1.000 UL
LTb
300 180 M
0 63 V
0 777 R
0 -63 V
1.000 UL
LTa
653 180 M
0 840 V
1.000 UL
LTb
653 180 M
0 63 V
0 777 R
0 -63 V
1.000 UL
LTa
1005 180 M
0 840 V
1.000 UL
LTb
1005 180 M
0 63 V
0 777 R
0 -63 V
1.000 UL
LTa
1358 180 M
0 840 V
1.000 UL
LTb
1358 180 M
0 63 V
0 777 R
0 -63 V
1.000 UL
LTa
1710 180 M
0 840 V
1.000 UL
LTb
1710 180 M
0 63 V
0 777 R
0 -63 V
1.000 UL
LTa
300 336 M
1410 0 V
1.000 UL
LTa
1005 180 M
0 840 V
1.000 UL
LTb
300 180 M
1410 0 V
0 840 V
-1410 0 V
0 -840 V
1.000 UL
LT0
300 355 M
28 0 V
28 1 V
29 1 V
28 2 V
28 2 V
28 3 V
28 3 V
29 5 V
28 6 V
28 6 V
28 8 V
28 10 V
29 11 V
28 13 V
28 14 V
28 16 V
28 16 V
29 18 V
28 17 V
28 17 V
28 16 V
28 13 V
29 10 V
28 7 V
28 2 V
28 -2 V
28 -7 V
29 -10 V
28 -13 V
28 -16 V
28 -17 V
28 -17 V
29 -18 V
28 -16 V
28 -16 V
28 -14 V
28 -13 V
29 -11 V
28 -10 V
28 -8 V
28 -6 V
28 -6 V
29 -5 V
28 -3 V
28 -3 V
28 -2 V
28 -2 V
29 -1 V
28 -1 V
28 0 V
1.000 UL
LT1
300 334 M
28 0 V
28 0 V
29 0 V
28 0 V
28 1 V
28 0 V
28 1 V
29 2 V
28 3 V
28 3 V
28 6 V
28 7 V
29 11 V
28 14 V
28 18 V
28 24 V
28 30 V
29 35 V
28 40 V
28 42 V
28 43 V
28 38 V
29 31 V
28 21 V
28 7 V
28 -7 V
28 -21 V
29 -31 V
28 -38 V
28 -43 V
28 -42 V
28 -40 V
29 -35 V
28 -30 V
28 -24 V
28 -18 V
28 -14 V
29 -11 V
28 -7 V
28 -6 V
28 -3 V
28 -3 V
29 -2 V
28 -1 V
28 0 V
28 -1 V
28 0 V
29 0 V
28 0 V
28 0 V
1.000 UL
LT2
300 315 M
14 0 V
14 -1 V
15 0 V
14 -1 V
14 -1 V
14 -1 V
15 -1 V
14 -2 V
14 -2 V
14 -2 V
15 -2 V
14 -3 V
14 -3 V
14 -4 V
15 -3 V
14 -5 V
14 -4 V
14 -4 V
15 -5 V
14 -4 V
14 -4 V
14 -4 V
15 -2 V
14 -1 V
14 0 V
14 3 V
15 6 V
14 8 V
14 12 V
14 15 V
15 20 V
14 24 V
14 29 V
14 33 V
14 37 V
15 40 V
14 44 V
14 47 V
14 49 V
15 49 V
14 49 V
14 47 V
14 45 V
15 41 V
14 36 V
14 30 V
14 24 V
15 16 V
14 8 V
14 0 V
14 -8 V
15 -16 V
14 -24 V
14 -30 V
14 -36 V
15 -41 V
14 -45 V
14 -47 V
14 -49 V
15 -49 V
14 -49 V
14 -47 V
14 -44 V
15 -40 V
14 -37 V
14 -33 V
14 -29 V
14 -24 V
15 -20 V
14 -15 V
14 -12 V
14 -8 V
15 -6 V
14 -3 V
14 0 V
14 1 V
15 2 V
14 4 V
14 4 V
14 4 V
15 5 V
14 4 V
14 4 V
14 5 V
15 3 V
14 4 V
14 3 V
14 3 V
15 2 V
14 2 V
14 2 V
14 2 V
15 1 V
14 1 V
14 1 V
14 1 V
15 0 V
14 1 V
14 0 V
stroke
grestore
end
showpage
}}%
\put(1005,30){\makebox(0,0){$\phi_+-\phi^\prime_+$}}%
\put(1005,-150){\makebox(0,0){\bf (b)}}%
\put(60,600){%
\special{ps: gsave currentpoint currentpoint translate
270 rotate neg exch neg exch translate}%
\makebox(0,0)[b]{\shortstack{${\cal P}^{(+)}(\phi_+-\phi^\prime_+)$}}%
\special{ps: currentpoint grestore moveto}%
}%
\put(1710,120){\makebox(0,0){$\pi$}}%
\put(1358,120){\makebox(0,0){$\pi/2$}}%
\put(1005,120){\makebox(0,0){0}}%
\put(653,120){\makebox(0,0){$-\pi/2$}}%
\put(300,120){\makebox(0,0){$-\pi$}}%
\put(270,958){\makebox(0,0)[r]{1}}%
\put(270,802){\makebox(0,0)[r]{0.75}}%
\put(270,647){\makebox(0,0)[r]{0.5}}%
\put(270,491){\makebox(0,0)[r]{0.25}}%
\put(270,336){\makebox(0,0)[r]{0}}%
\put(270,180){\makebox(0,0)[r]{-0.25}}%
\end{picture}%
\endgroup
 

%% file: pw_sum_odd.tex
\begingroup%
  \makeatletter%
  \newcommand{\GNUPLOTspecial}{%
    \@sanitize\catcode`\%=14\relax\special}%
  \setlength{\unitlength}{0.1bp}%
\begin{picture}(3600,2160)(0,-500)%
\special{psfile=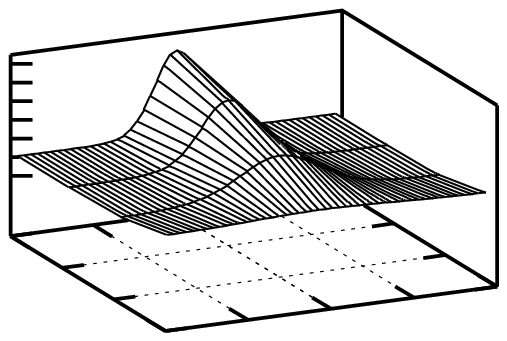 llx=0 lly=0 urx=720 ury=504 rwi=7200}
\put(1099,1574){\makebox(0,0){${\cal P}^{(+)}(\phi_+-\phi^\prime_+)$}}%
\put(1033,1417){\makebox(0,0)[r]{1}}%
\put(1033,1363){\makebox(0,0)[r]{0.8}}%
\put(1033,1310){\makebox(0,0)[r]{0.6}}%
\put(1033,1256){\makebox(0,0)[r]{0.4}}%
\put(1033,1202){\makebox(0,0)[r]{0.2}}%
\put(1033,1149){\makebox(0,0)[r]{0}}%
\put(1033,1095){\makebox(0,0)[r]{-0.2}}%
\put(2134,560){\makebox(0,0){$\phi_+-\phi^\prime_+$}}%
\put(2134,400){\makebox(0,0){\bf (c)}}%
\put(1570,596){\makebox(0,0){$-\pi$}}%
\put(1808,628){\makebox(0,0){$-\pi/2$}}%
\put(2047,660){\makebox(0,0){0}}%
\put(2286,692){\makebox(0,0){$\pi/2$}}%
\put(2525,724){\makebox(0,0){$\pi$}}%
\put(1196,684){\makebox(0,0){$|\al|^2$}}%
\put(1069,903){\makebox(0,0)[r]{3}}%
\put(1218,812){\makebox(0,0)[r]{2}}%
\put(1366,721){\makebox(0,0)[r]{1}}%
\put(1515,630){\makebox(0,0)[r]{0}}%
\end{picture}%
\endgroup
 

%% file: pp_sum_odd.tex
\begingroup%
  \makeatletter%
  \newcommand{\GNUPLOTspecial}{%
    \@sanitize\catcode`\%=14\relax\special}%
  \setlength{\unitlength}{0.1bp}%
{\GNUPLOTspecial{!
/gnudict 256 dict def
gnudict begin
/Color false def
/Solid false def
/gnulinewidth 5.000 def
/userlinewidth gnulinewidth def
/vshift -20 def
/dl {10 mul} def
/hpt_ 31.5 def
/vpt_ 31.5 def
/hpt hpt_ def
/vpt vpt_ def
/M {moveto} bind def
/L {lineto} bind def
/R {rmoveto} bind def
/V {rlineto} bind def
/vpt2 vpt 2 mul def
/hpt2 hpt 2 mul def
/Lshow { currentpoint stroke M
  0 vshift R show } def
/Rshow { currentpoint stroke M
  dup stringwidth pop neg vshift R show } def
/Cshow { currentpoint stroke M
  dup stringwidth pop -2 div vshift R show } def
/UP { dup vpt_ mul /vpt exch def hpt_ mul /hpt exch def
  /hpt2 hpt 2 mul def /vpt2 vpt 2 mul def } def
/DL { Color {setrgbcolor Solid {pop []} if 0 setdash }
 {pop pop pop Solid {pop []} if 0 setdash} ifelse } def
/BL { stroke userlinewidth 2 mul setlinewidth } def
/AL { stroke userlinewidth 2 div setlinewidth } def
/UL { dup gnulinewidth mul /userlinewidth exch def
      10 mul /udl exch def } def
/PL { stroke userlinewidth setlinewidth } def
/LTb { BL [] 0 0 0 DL } def
/LTa { AL [1 udl mul 2 udl mul] 0 setdash 0 0 0 setrgbcolor } def
/LT0 { PL [] 1 0 0 DL } def
/LT1 { PL [4 dl 2 dl] 0 1 0 DL } def
/LT2 { PL [2 dl 3 dl] 0 0 1 DL } def
/LT3 { PL [1 dl 1.5 dl] 1 0 1 DL } def
/LT4 { PL [5 dl 2 dl 1 dl 2 dl] 0 1 1 DL } def
/LT5 { PL [4 dl 3 dl 1 dl 3 dl] 1 1 0 DL } def
/LT6 { PL [2 dl 2 dl 2 dl 4 dl] 0 0 0 DL } def
/LT7 { PL [2 dl 2 dl 2 dl 2 dl 2 dl 4 dl] 1 0.3 0 DL } def
/LT8 { PL [2 dl 2 dl 2 dl 2 dl 2 dl 2 dl 2 dl 4 dl] 0.5 0.5 0.5 DL } def
/Pnt { stroke [] 0 setdash
   gsave 1 setlinecap M 0 0 V stroke grestore } def
/Dia { stroke [] 0 setdash 2 copy vpt add M
  hpt neg vpt neg V hpt vpt neg V
  hpt vpt V hpt neg vpt V closepath stroke
  Pnt } def
/Pls { stroke [] 0 setdash vpt sub M 0 vpt2 V
  currentpoint stroke M
  hpt neg vpt neg R hpt2 0 V stroke
  } def
/Box { stroke [] 0 setdash 2 copy exch hpt sub exch vpt add M
  0 vpt2 neg V hpt2 0 V 0 vpt2 V
  hpt2 neg 0 V closepath stroke
  Pnt } def
/Crs { stroke [] 0 setdash exch hpt sub exch vpt add M
  hpt2 vpt2 neg V currentpoint stroke M
  hpt2 neg 0 R hpt2 vpt2 V stroke } def
/TriU { stroke [] 0 setdash 2 copy vpt 1.12 mul add M
  hpt neg vpt -1.62 mul V
  hpt 2 mul 0 V
  hpt neg vpt 1.62 mul V closepath stroke
  Pnt  } def
/Star { 2 copy Pls Crs } def
/BoxF { stroke [] 0 setdash exch hpt sub exch vpt add M
  0 vpt2 neg V  hpt2 0 V  0 vpt2 V
  hpt2 neg 0 V  closepath fill } def
/TriUF { stroke [] 0 setdash vpt 1.12 mul add M
  hpt neg vpt -1.62 mul V
  hpt 2 mul 0 V
  hpt neg vpt 1.62 mul V closepath fill } def
/TriD { stroke [] 0 setdash 2 copy vpt 1.12 mul sub M
  hpt neg vpt 1.62 mul V
  hpt 2 mul 0 V
  hpt neg vpt -1.62 mul V closepath stroke
  Pnt  } def
/TriDF { stroke [] 0 setdash vpt 1.12 mul sub M
  hpt neg vpt 1.62 mul V
  hpt 2 mul 0 V
  hpt neg vpt -1.62 mul V closepath fill} def
/DiaF { stroke [] 0 setdash vpt add M
  hpt neg vpt neg V hpt vpt neg V
  hpt vpt V hpt neg vpt V closepath fill } def
/Pent { stroke [] 0 setdash 2 copy gsave
  translate 0 hpt M 4 {72 rotate 0 hpt L} repeat
  closepath stroke grestore Pnt } def
/PentF { stroke [] 0 setdash gsave
  translate 0 hpt M 4 {72 rotate 0 hpt L} repeat
  closepath fill grestore } def
/Circle { stroke [] 0 setdash 2 copy
  hpt 0 360 arc stroke Pnt } def
/CircleF { stroke [] 0 setdash hpt 0 360 arc fill } def
/C0 { BL [] 0 setdash 2 copy moveto vpt 90 450  arc } bind def
/C1 { BL [] 0 setdash 2 copy        moveto
       2 copy  vpt 0 90 arc closepath fill
               vpt 0 360 arc closepath } bind def
/C2 { BL [] 0 setdash 2 copy moveto
       2 copy  vpt 90 180 arc closepath fill
               vpt 0 360 arc closepath } bind def
/C3 { BL [] 0 setdash 2 copy moveto
       2 copy  vpt 0 180 arc closepath fill
               vpt 0 360 arc closepath } bind def
/C4 { BL [] 0 setdash 2 copy moveto
       2 copy  vpt 180 270 arc closepath fill
               vpt 0 360 arc closepath } bind def
/C5 { BL [] 0 setdash 2 copy moveto
       2 copy  vpt 0 90 arc
       2 copy moveto
       2 copy  vpt 180 270 arc closepath fill
               vpt 0 360 arc } bind def
/C6 { BL [] 0 setdash 2 copy moveto
      2 copy  vpt 90 270 arc closepath fill
              vpt 0 360 arc closepath } bind def
/C7 { BL [] 0 setdash 2 copy moveto
      2 copy  vpt 0 270 arc closepath fill
              vpt 0 360 arc closepath } bind def
/C8 { BL [] 0 setdash 2 copy moveto
      2 copy vpt 270 360 arc closepath fill
              vpt 0 360 arc closepath } bind def
/C9 { BL [] 0 setdash 2 copy moveto
      2 copy  vpt 270 450 arc closepath fill
              vpt 0 360 arc closepath } bind def
/C10 { BL [] 0 setdash 2 copy 2 copy moveto vpt 270 360 arc closepath fill
       2 copy moveto
       2 copy vpt 90 180 arc closepath fill
               vpt 0 360 arc closepath } bind def
/C11 { BL [] 0 setdash 2 copy moveto
       2 copy  vpt 0 180 arc closepath fill
       2 copy moveto
       2 copy  vpt 270 360 arc closepath fill
               vpt 0 360 arc closepath } bind def
/C12 { BL [] 0 setdash 2 copy moveto
       2 copy  vpt 180 360 arc closepath fill
               vpt 0 360 arc closepath } bind def
/C13 { BL [] 0 setdash  2 copy moveto
       2 copy  vpt 0 90 arc closepath fill
       2 copy moveto
       2 copy  vpt 180 360 arc closepath fill
               vpt 0 360 arc closepath } bind def
/C14 { BL [] 0 setdash 2 copy moveto
       2 copy  vpt 90 360 arc closepath fill
               vpt 0 360 arc } bind def
/C15 { BL [] 0 setdash 2 copy vpt 0 360 arc closepath fill
               vpt 0 360 arc closepath } bind def
/Rec   { newpath 4 2 roll moveto 1 index 0 rlineto 0 exch rlineto
       neg 0 rlineto closepath } bind def
/Square { dup Rec } bind def
/Bsquare { vpt sub exch vpt sub exch vpt2 Square } bind def
/S0 { BL [] 0 setdash 2 copy moveto 0 vpt rlineto BL Bsquare } bind def
/S1 { BL [] 0 setdash 2 copy vpt Square fill Bsquare } bind def
/S2 { BL [] 0 setdash 2 copy exch vpt sub exch vpt Square fill Bsquare } bind def
/S3 { BL [] 0 setdash 2 copy exch vpt sub exch vpt2 vpt Rec fill Bsquare } bind def
/S4 { BL [] 0 setdash 2 copy exch vpt sub exch vpt sub vpt Square fill Bsquare } bind def
/S5 { BL [] 0 setdash 2 copy 2 copy vpt Square fill
       exch vpt sub exch vpt sub vpt Square fill Bsquare } bind def
/S6 { BL [] 0 setdash 2 copy exch vpt sub exch vpt sub vpt vpt2 Rec fill Bsquare } bind def
/S7 { BL [] 0 setdash 2 copy exch vpt sub exch vpt sub vpt vpt2 Rec fill
       2 copy vpt Square fill
       Bsquare } bind def
/S8 { BL [] 0 setdash 2 copy vpt sub vpt Square fill Bsquare } bind def
/S9 { BL [] 0 setdash 2 copy vpt sub vpt vpt2 Rec fill Bsquare } bind def
/S10 { BL [] 0 setdash 2 copy vpt sub vpt Square fill 2 copy exch vpt sub exch vpt Square fill
       Bsquare } bind def
/S11 { BL [] 0 setdash 2 copy vpt sub vpt Square fill 2 copy exch vpt sub exch vpt2 vpt Rec fill
       Bsquare } bind def
/S12 { BL [] 0 setdash 2 copy exch vpt sub exch vpt sub vpt2 vpt Rec fill Bsquare } bind def
/S13 { BL [] 0 setdash 2 copy exch vpt sub exch vpt sub vpt2 vpt Rec fill
       2 copy vpt Square fill Bsquare } bind def
/S14 { BL [] 0 setdash 2 copy exch vpt sub exch vpt sub vpt2 vpt Rec fill
       2 copy exch vpt sub exch vpt Square fill Bsquare } bind def
/S15 { BL [] 0 setdash 2 copy Bsquare fill Bsquare } bind def
/D0 { gsave translate 45 rotate 0 0 S0 stroke grestore } bind def
/D1 { gsave translate 45 rotate 0 0 S1 stroke grestore } bind def
/D2 { gsave translate 45 rotate 0 0 S2 stroke grestore } bind def
/D3 { gsave translate 45 rotate 0 0 S3 stroke grestore } bind def
/D4 { gsave translate 45 rotate 0 0 S4 stroke grestore } bind def
/D5 { gsave translate 45 rotate 0 0 S5 stroke grestore } bind def
/D6 { gsave translate 45 rotate 0 0 S6 stroke grestore } bind def
/D7 { gsave translate 45 rotate 0 0 S7 stroke grestore } bind def
/D8 { gsave translate 45 rotate 0 0 S8 stroke grestore } bind def
/D9 { gsave translate 45 rotate 0 0 S9 stroke grestore } bind def
/D10 { gsave translate 45 rotate 0 0 S10 stroke grestore } bind def
/D11 { gsave translate 45 rotate 0 0 S11 stroke grestore } bind def
/D12 { gsave translate 45 rotate 0 0 S12 stroke grestore } bind def
/D13 { gsave translate 45 rotate 0 0 S13 stroke grestore } bind def
/D14 { gsave translate 45 rotate 0 0 S14 stroke grestore } bind def
/D15 { gsave translate 45 rotate 0 0 S15 stroke grestore } bind def
/DiaE { stroke [] 0 setdash vpt add M
  hpt neg vpt neg V hpt vpt neg V
  hpt vpt V hpt neg vpt V closepath stroke } def
/BoxE { stroke [] 0 setdash exch hpt sub exch vpt add M
  0 vpt2 neg V hpt2 0 V 0 vpt2 V
  hpt2 neg 0 V closepath stroke } def
/TriUE { stroke [] 0 setdash vpt 1.12 mul add M
  hpt neg vpt -1.62 mul V
  hpt 2 mul 0 V
  hpt neg vpt 1.62 mul V closepath stroke } def
/TriDE { stroke [] 0 setdash vpt 1.12 mul sub M
  hpt neg vpt 1.62 mul V
  hpt 2 mul 0 V
  hpt neg vpt -1.62 mul V closepath stroke } def
/PentE { stroke [] 0 setdash gsave
  translate 0 hpt M 4 {72 rotate 0 hpt L} repeat
  closepath stroke grestore } def
/CircE { stroke [] 0 setdash 
  hpt 0 360 arc stroke } def
/Opaque { gsave closepath 1 setgray fill grestore 0 setgray closepath } def
/DiaW { stroke [] 0 setdash vpt add M
  hpt neg vpt neg V hpt vpt neg V
  hpt vpt V hpt neg vpt V Opaque stroke } def
/BoxW { stroke [] 0 setdash exch hpt sub exch vpt add M
  0 vpt2 neg V hpt2 0 V 0 vpt2 V
  hpt2 neg 0 V Opaque stroke } def
/TriUW { stroke [] 0 setdash vpt 1.12 mul add M
  hpt neg vpt -1.62 mul V
  hpt 2 mul 0 V
  hpt neg vpt 1.62 mul V Opaque stroke } def
/TriDW { stroke [] 0 setdash vpt 1.12 mul sub M
  hpt neg vpt 1.62 mul V
  hpt 2 mul 0 V
  hpt neg vpt -1.62 mul V Opaque stroke } def
/PentW { stroke [] 0 setdash gsave
  translate 0 hpt M 4 {72 rotate 0 hpt L} repeat
  Opaque stroke grestore } def
/CircW { stroke [] 0 setdash 
  hpt 0 360 arc Opaque stroke } def
/BoxFill { gsave Rec 1 setgray fill grestore } def
/Symbol-Oblique /Symbol findfont [1 0 .167 1 0 0] makefont
dup length dict begin {1 index /FID eq {pop pop} {def} ifelse} forall
currentdict end definefont
end
}}%
\begin{picture}(1800,1080)(0,-1040)%
{\GNUPLOTspecial{"
gnudict begin
gsave
0 0 translate
0.100 0.100 scale
0 setgray
newpath
1.000 UL
LTb
1.000 UL
LTa
300 180 M
1410 0 V
1.000 UL
LTb
300 180 M
63 0 V
1347 0 R
-63 0 V
1.000 UL
LTa
300 460 M
1410 0 V
1.000 UL
LTb
300 460 M
63 0 V
1347 0 R
-63 0 V
1.000 UL
LTa
300 740 M
1410 0 V
1.000 UL
LTb
300 740 M
63 0 V
1347 0 R
-63 0 V
1.000 UL
LTa
300 1020 M
1410 0 V
1.000 UL
LTb
300 1020 M
63 0 V
1347 0 R
-63 0 V
1.000 UL
LTa
300 180 M
0 840 V
1.000 UL
LTb
300 180 M
0 63 V
0 777 R
0 -63 V
1.000 UL
LTa
653 180 M
0 840 V
1.000 UL
LTb
653 180 M
0 63 V
0 777 R
0 -63 V
1.000 UL
LTa
1005 180 M
0 840 V
1.000 UL
LTb
1005 180 M
0 63 V
0 777 R
0 -63 V
1.000 UL
LTa
1358 180 M
0 840 V
1.000 UL
LTb
1358 180 M
0 63 V
0 777 R
0 -63 V
1.000 UL
LTa
1710 180 M
0 840 V
1.000 UL
LTb
1710 180 M
0 63 V
0 777 R
0 -63 V
1.000 UL
LTa
300 460 M
1410 0 V
1.000 UL
LTa
1005 180 M
0 840 V
1.000 UL
LTb
300 180 M
1410 0 V
0 840 V
-1410 0 V
0 -840 V
1.000 UL
LT0
300 503 M
28 0 V
28 1 V
29 2 V
28 3 V
28 4 V
28 5 V
28 7 V
29 8 V
28 10 V
28 11 V
28 14 V
28 16 V
29 19 V
28 22 V
28 24 V
28 26 V
28 28 V
29 29 V
28 30 V
28 28 V
28 25 V
28 22 V
29 16 V
28 11 V
28 3 V
28 -3 V
28 -11 V
29 -16 V
28 -22 V
28 -25 V
28 -28 V
28 -30 V
29 -29 V
28 -28 V
28 -26 V
28 -24 V
28 -22 V
29 -19 V
28 -16 V
28 -14 V
28 -11 V
28 -10 V
29 -8 V
28 -7 V
28 -5 V
28 -4 V
28 -3 V
29 -2 V
28 -1 V
28 0 V
1.000 UL
LT1
300 483 M
28 0 V
28 1 V
29 1 V
28 2 V
28 3 V
28 3 V
28 5 V
29 6 V
28 7 V
28 10 V
28 12 V
28 16 V
29 20 V
28 23 V
28 29 V
28 34 V
28 39 V
29 43 V
28 47 V
28 47 V
28 45 V
28 40 V
29 31 V
28 20 V
28 7 V
28 -7 V
28 -20 V
29 -31 V
28 -40 V
28 -45 V
28 -47 V
28 -47 V
29 -43 V
28 -39 V
28 -34 V
28 -29 V
28 -23 V
29 -20 V
28 -16 V
28 -12 V
28 -10 V
28 -7 V
29 -6 V
28 -5 V
28 -3 V
28 -3 V
28 -2 V
29 -1 V
28 -1 V
28 0 V
1.000 UL
LT2
300 502 M
14 0 V
14 1 V
15 0 V
14 1 V
14 2 V
14 1 V
15 3 V
14 2 V
14 3 V
14 4 V
15 4 V
14 5 V
14 5 V
14 7 V
15 7 V
14 9 V
14 10 V
14 11 V
15 14 V
14 14 V
14 17 V
14 20 V
15 21 V
14 23 V
14 26 V
14 28 V
15 30 V
14 31 V
14 31 V
14 32 V
15 29 V
14 27 V
14 21 V
14 15 V
14 7 V
15 -4 V
14 -16 V
14 -28 V
14 -42 V
15 -54 V
14 -66 V
14 -75 V
14 -80 V
15 -82 V
14 -78 V
14 -70 V
14 -57 V
15 -41 V
14 -21 V
14 0 V
14 21 V
15 41 V
14 57 V
14 70 V
14 78 V
15 82 V
14 80 V
14 75 V
14 66 V
15 54 V
14 42 V
14 28 V
14 16 V
15 4 V
14 -7 V
14 -15 V
14 -21 V
14 -27 V
15 -29 V
14 -32 V
14 -31 V
14 -31 V
15 -30 V
14 -28 V
14 -26 V
14 -23 V
15 -21 V
14 -20 V
14 -17 V
14 -14 V
15 -14 V
14 -11 V
14 -10 V
14 -9 V
15 -7 V
14 -7 V
14 -5 V
14 -5 V
15 -4 V
14 -4 V
14 -3 V
14 -2 V
15 -3 V
14 -1 V
14 -2 V
14 -1 V
15 0 V
14 -1 V
14 0 V
stroke
grestore
end
showpage
}}%
\put(1005,30){\makebox(0,0){$\phi_+-\phi^\prime_+$}}%
\put(1005,-150){\makebox(0,0){\bf (d)}}%
\put(60,600){%
\special{ps: gsave currentpoint currentpoint translate
270 rotate neg exch neg exch translate}%
\makebox(0,0)[b]{\shortstack{${\cal P}^{(+)}(\phi_+-\phi^\prime_+)$}}%
\special{ps: currentpoint grestore moveto}%
}%
\put(1710,120){\makebox(0,0){$\pi$}}%
\put(1358,120){\makebox(0,0){$\pi/2$}}%
\put(1005,120){\makebox(0,0){0}}%
\put(653,120){\makebox(0,0){$-\pi/2$}}%
\put(300,120){\makebox(0,0){$-\pi$}}%
\put(270,1020){\makebox(0,0)[r]{0.5}}%
\put(270,740){\makebox(0,0)[r]{0.25}}%
\put(270,460){\makebox(0,0)[r]{0}}%
\put(270,180){\makebox(0,0)[r]{-0.25}}%
\end{picture}%
\endgroup
 